\documentclass[12pt,preprint]{aastex}
\usepackage{lscape}
\usepackage{amsmath}
\usepackage{booktabs}

\def\ba{\begin{eqnarray}}
\def\ea{\end{eqnarray}}

\begin{document}
\title{The missing cavities in the SEEDS polarized scattered light images of transitional protoplanetary disks: a generic disk model}

\shorttitle{The missing cavities}

\shortauthors{Dong et al.}

\author{R. Dong\altaffilmark{1}, R. Rafikov\altaffilmark{1}, Z. Zhu\altaffilmark{1}, L. Hartmann\altaffilmark{2}, B. Whitney\altaffilmark{3}, T. Brandt\altaffilmark{1}, T. Muto\altaffilmark{4,5}, J. Hashimoto\altaffilmark{6}, C. Grady\altaffilmark{7,8}, K. Follette\altaffilmark{9}, M. Kuzuhara\altaffilmark{6}, R. Tanii\altaffilmark{10}, Y. Itoh\altaffilmark{10}, C. Thalmann\altaffilmark{11}, J. Wisniewski\altaffilmark{12}, S. Mayama\altaffilmark{13}, M. Janson\altaffilmark{1}, L. Abe\altaffilmark{14}, W. Brandner\altaffilmark{15}, J. Carson\altaffilmark{16}, S. Egner\altaffilmark{17}, M. Feldt\altaffilmark{15}, M. Goto\altaffilmark{15}, O. Guyon\altaffilmark{17}, Y. Hayano\altaffilmark{17}, M. Hayashi\altaffilmark{18}, S. Hayashi\altaffilmark{17}, T. Henning\altaffilmark{15}, K. W. Hodapp\altaffilmark{19}, M. Honda\altaffilmark{20}, S. Inutsuka\altaffilmark{21}, M. Ishii\altaffilmark{17}, M. Iye\altaffilmark{6}, R. Kandori\altaffilmark{6}, G. R. Knapp\altaffilmark{1}, T. Kudo\altaffilmark{17}, N. Kusakabe\altaffilmark{6}, T. Matsuo\altaffilmark{22}, M. W. McElwain\altaffilmark{7}, S. Miyama\altaffilmark{6}, J.-I. Morino\altaffilmark{6}, A. Moro-Martin\altaffilmark{23}, T. Nishimura\altaffilmark{17}, T.-S. Pyo\altaffilmark{17}, H. Suto\altaffilmark{6}, R. Suzuki\altaffilmark{24}, M. Takami\altaffilmark{25}, N. Takato\altaffilmark{17}, H. Terada\altaffilmark{17}, D. Tomono\altaffilmark{17}, E. L. Turner\altaffilmark{1,26}, M. Watanabe\altaffilmark{27}, T. Yamada\altaffilmark{28}, H. Takami\altaffilmark{17}, T. Usuda\altaffilmark{17}, M. Tamura\altaffilmark{6}}

\altaffiltext{1}{Department of Astrophysical Sciences, Princeton University, Princeton, NJ, 08544, USA; rdong@astro.princeton.edu}
\altaffiltext{2}{Department of Astronomy, University of Michigan, 500 Church Street, Ann Arbor, MI 48105, USA}
\altaffiltext{3}{Astronomy Department, University of Wisconsin-Madison, 475 N. Charter St., Madison, WI 53706, USA}
\altaffiltext{4}{Tokyo Institute of Technology, 2-12-1 Ookayama, Meguro, Tokyo 152-8551, Japan}
\altaffiltext{5}{Division of Liberal Arts, Kogakuin University, 1-24-2, Nishi-Shinjuku, Shinjuku-ku, Tokyo, 163-8677, Japan}
\altaffiltext{6}{National Astronomical Observatory of Japan, 2-21-1 Osawa, Mitaka, Tokyo 181-8588, Japan}
\altaffiltext{7}{ExoPlanets and Stellar Astrophysics Laboratory, Code 667, Goddard Space Flight Center, Greenbelt, MD 20771 USA}
\altaffiltext{8}{Eureka Scientific, 2452 Delmer, Suite 100, Oakland CA 96002, USA}
\altaffiltext{9}{Steward Observatory, University of Arizona, 933 North Cherry Avenue, Tucson, AZ 85721, USA}
\altaffiltext{10}{Graduate School of Science, Kobe University, 1-1 Rokkodai, Nada-ku, Kobe 657-8501, Japan}
\altaffiltext{11}{Astronomical Institute "Anton Pannekoek", University of Amsterdam, Science Park 904, 1098 XH Amsterdam, The Netherlands}
\altaffiltext{12}{Department of Astronomy, University of Washington, Box 351580 Seattle, Washington 98195, USA}
\altaffiltext{13}{The Graduate University for Advanced Studies(SOKENDAI), Shonan International Village, Hayama-cho, Miura-gun, Kanagawa 240-0193, Japan}
\altaffiltext{14} {Laboratoire Lagrange, UMR7293, Universit\'e de Nice-Sophia Antipolis, 
CNRS, Observatoire de la C\^ote d'Azur, 06300 Nice, France}
\altaffiltext{15} {Max Planck Institute for Astronomy, Heidelberg, Germany}
\altaffiltext{16} {Department of Physics and Astronomy, College of
Charleston, 58 Coming St., Charleston, SC 29424, USA}
\altaffiltext{17} {Subaru Telescope, 650 North A'ohoku Place, Hilo, HI
96720, USA}
\altaffiltext{18} {Department of Astronomy, The University of Tokyo,
Hongo 7-3-1, Bunkyo-ku, Tokyo 113-0033, Japan}
\altaffiltext{19} {Institute for Astronomy, University of Hawaii, 640
North A'ohoku Place, Hilo, HI 96720, USA}
\altaffiltext{20} {Department of Information Sciences, Kanagawa University, 2946 Tsuchiya, Hiratsuka, Kanagawa 259-1293, Japan}
\altaffiltext{21} {Department of Physics, Graduate School of Science, Nagoya University, Furo-cho, Chikusa-ku, Nagoya, Aichi 464-8602, Japan}
\altaffiltext{22} {Department of Astronomy, Kyoto University,
Kitashirakawa-Oiwake-cho, Sakyo-ku, Kyoto, 606-8502, Japan}
\altaffiltext{23} {Departamento de Astrof\'isica, CAB (INTA-CSIC), 
Instituto Nacional de T\'ecnica Aeroespacial, Torrej\'on de Ardoz,
28850, Madrid, Spain}
\altaffiltext{24} {TMT Observatory Corporation, 1111 South Arroyo
Parkway, Pasadena, CA 91105, USA}
\altaffiltext {25} {Institute of Astronomy and Astrophysics, Academia
Sinica, P.O. Box 23-141, Taipei 106, Taiwan}
\altaffiltext{26} {Institute for the Physics and Mathematics of the
Universe, The University of Tokyo, Kashiwa 227-8568, Japan}
\altaffiltext{27} {Department of Cosmosciences, Hokkaido University,
Sapporo 060-0810, Japan}
\altaffiltext{28} {Astronomical Institute, Tohoku University, Aoba,
Sendai 980-8578, Japan}

\begin{abstract}

Transitional circumstellar disks around young stellar objects have a distinctive infrared deficit around 10 microns in their Spectral Energy Distributions (SED), recently measured by the {\it Spitzer Infrared Spectrograph} (IRS), suggesting dust depletion in the inner regions. These disks have been confirmed to have giant central cavities by imaging of the submillimeter (sub-mm) continuum emission using the {\it Submillimeter Array} (SMA). However, the polarized near-infrared scattered light images for most objects in a systematic IRS/SMA cross sample, obtained by HiCIAO on the Subaru telescope, show no evidence for the cavity, in clear contrast with SMA and Spitzer observations. Radiative transfer modeling indicates that many of these scattered light images are consistent with a smooth spatial distribution for micron-sized grains, with little discontinuity in the surface density of the micron-sized grains at the cavity edge. Here we present a generic disk model that can simultaneously account for the general features in IRS, SMA, and Subaru observations. Particularly, the scattered light images for this model are computed, which agree with the general trend seen in Subaru data. Decoupling between the spatial distributions of the micron-sized dust and mm-sized dust inside the cavity is suggested by the model, which, if confirmed, necessitates a mechanism, such as dust filtration, for differentiating the small and big dust in the cavity clearing process. Our model also suggests an inwardly increasing gas-to-dust-ratio in the inner disk, and different spatial distributions for the small dust inside and outside the cavity, echoing the predictions in grain coagulation and growth models.

\end{abstract}


%%%%%%%%%%%%%%%%%%%%%%%%%%%%%%%%%%%%%%%%%%%%%%%%%%%%

\section{Introduction}\label{sec:introduction}

Transitional disks have a small or no excess from $\sim$$1\mu$m to $\sim$$10\mu$m relative to their full disk cousins, but a significant excess at longer wavelength \citep{wil11}, suggesting cleared out inner disk. This interpretation dates back to the era of the {\it Infrared Astronomical Satellite} (\citealt{str89}, \citealt{skr90}), and was later developed with the help of detailed near-infrared (NIR) to mid-infrared (MIR) spectra provided by the Infrared Spectrograph (IRS) on-board {\it Spitzer Space Telescope}. Detailed radiative transfer modeling suggests that this kind of SED is consistent with disk models which harbor a central (partially) depleted region (i.e. a cavity or a gap), while a ``wall-like'' structure at the outer edge of this region can be responsible for the abrupt rise of the SED at MIR \citep{cal05,esp07}\footnote{The objects with a small dust belt left at the center have been prototyped as pre-transitional objects, such as LkCa 15 \citep{esp08}, but for simplicity we use the term transitional disk for both types.}.

The disk+cavity model based on SED-only fitting usually contains large uncertainties, because the SED samples the emission from the whole disk; by tuning the ingredients in the fitting, one could fit the IRS SED with different models (see the example of UX Tau A, \citealt{esp10, esp11} and \citealt[hereafter A11]{and11}). Better constraints on the disk structure can be obtained from resolved images of the transitional disks. Using the {\it Submillimeter Array} (SMA) interferometer \citep{ho04}, resolved images of transitional disks at sub-mm wavelength have provided direct detections of these cavities \citep{and07,and09}, and measurements of their properties. Recently, A11 observed a sample of 12 nearby transitional disks (at a typical distance of $\sim$$140$~pc). Combining both the SMA results and the SED, they fit detailed disk+cavity model for each object, and the cavity size ($\sim$$15-70$~AU) is determined with $\sim$$10\%$ uncertainty. They concluded that large grains (up to $\sim$mm-sized) inside the cavity are depleted by at least a factor of 10 to 100 (the ``depletion'' in this work is relative to a ``background'' value extrapolated from the outer disk). Under the assumption that the surface density of the disk is described by their model, the infrared spectral fitting demands that the small grains (micron-sized and smaller) inside the cavity to be heavily depleted by a factor of $\sim$$10^5$.

Recently, most objects in this sample have been observed by the Subaru High-Contrast Coronographic Imager for Adaptive Optics (HiCIAO) at NIR bands, as part of the the Strategic Explorations of Exoplanets and Disks with Subaru project, SEEDS, \citep{tam09}. SEEDS is capable of producing polarized intensity (PI) images of disks, which greatly enhances our ability to probe disk structure (especially at the inner part) by utilizing the fact that the central source is usually not polarized, so that the stellar residual in PI images is much smaller than in full intensity (FI) images \citep{per04,hin09,qua11}.

The SEEDS results turned out to be a big surprise --- in many cases the polarized NIR images do not show an inner cavity, despite the fact that the inner working angle of the images (the saturation radius or the coronagraph mask size, $\psi_{\rm in} \sim 0.\!\!''1-0.\!\!''15$, or $\sim$$15-25$~AU at the distance to Taurus $\sim$$140$~pc, see Section~\ref{sec:psf}) is significantly smaller than the cavity sizes inferred from sub-mm observations. High contrast features such as surface brightness excesses or deficits exist in some systems, but they are localized and do not appear to be central cavities. Instead, the image is smooth on large scales, and the azimuthally averaged surface brightness radial profile (or the profile along the major axis) increases inward smoothly until $\psi_{\rm in}$, without any abrupt break or jump at the cavity edge (the slope may change with radius in some systems). Examples include ROX 44 (M. Kuzuhara et al. 2012, in prep.), SR 21 (K. Follette et al. 2012, in prep.), GM Aur (J. Hashimoto et al. 2012a, in prep.), and SAO 206462 \citep{mut12}; see also the sample statistics (J. Hashimoto et al. 2012b, in prep.). Some objects such as UX Tau A also do not show a cavity (R. Tanii et al. 2012, in prep.), however the inner working angle of their SEEDS images is too close to the cavity size, so the status of the cavity is less certain. We note that LkCa 15 also does not exhibit a clear cavity in its PI imagery (J. Wisniewski et al. 2012, in prep.), but does exhibit evidence of the wall of a cavity in its FI imagery \citep{tha10}.

This apparent inconsistency between observations at different wavelengths reveals something fundamental in the transitional disk structure, as these datasets probe different components of protoplanetary disks. At short wavelengths (i.e. NIR) where the disk is optically thick, the flux is dominated by the small dust (micron-sized or so) at the surface of the disk (where the stellar photons get absorbed or scattered), and is sensitive to the shape of the surface; at long wavelength (i.e. sub-mm), disks are generally optically thin, so the flux essentially probes the disk surface density in big grains (mm-sized or so), due to their large opacity at these wavelengths \citep{wil11}.

Combining all the three pieces of the puzzle together (SED, sub-mm observation, and NIR imaging), we propose a disk model that explains the signatures in all three observations simultaneously: the key point is that the spatial distributions of small and big dust are decoupled inside the cavity. In this model, a well defined cavity (several tens of~AU in radius) with a sharp edge exists only in spatial distribution of the big dust and reproduces the central void in the sub-mm images, while no discontinuity is found for the spatial distribution of the small dust at the cavity edge. Inside the cavity, the surface density of the small dust does not increase inwardly as steeply as it does in the outer disk; instead it is roughly constant or declines closer to the star (while maintaining an overall smooth profile). In this way, the inner region (sub-AU to a few~AU) is heavily depleted in small dust, so that the model reproduces the NIR flux deficit in the SED (but still enough small dust surface density to efficiently scatter near-IR radiation). Modeling results show that the scattered light images for this continuous spatial distribution of the small dust appear smooth as well, with surface brightness steadily increasing inwardly, as seen in many of the SEEDS observations.

The structure of this paper is as follows. In Section~\ref{sec:modeling} we introduce the method that we use for the radiative transfer modeling. In Section~\ref{sec:image} we give the main results on the scattered light images: first a general interpretation of the big picture through a theoretical perspective, followed by the modeling results of various disk+cavity models. We investigate the sub-mm properties of these models in Section~\ref{sec:submm}, and explore the degeneracy in the disk parameter space on their model SED in Section~\ref{sec:sed}. We summarize the direct constraints put by the three observations on this transitional disk sample in Section~\ref{sec:discussion}, as well as the implications of our disk models. Our generic solution, which qualitatively explains the signatures in all the three observations, is summarized in Section~\ref{sec:summary}.

%%%%%%%%%%%%%%%%%%%%%%%%%%%%%%%%%%%%%%%%%%%%%%%%%%%%

%%%%%%%%%%%%%%%%%%%%%%%%%%%%%%%%%%%%%%%%%%%%%%%%%%%%

\section{Radiative Transfer Modeling}\label{sec:modeling}

In this section, we introduce the model setup in our radiative transfer calculations, and the post processing of the raw NIR polarized scattered light images which we perform in order to mimic the observations. The purpose of this modeling exercise is to ``translate'' various physical disk models to their corresponding NIR polarized scattered light images, sub-mm emission images, and SED, for comparison with observations.

\subsection{Model setup}\label{sec:setup}

We use a modified version of the Monte Carlo radiative transfer code developed by \citet{whi03a,whi03b}, \citet{rob06}, and B. Whitney et al. 2012, in prep.; for the disk structure, we use A11 and \citet{whi03b} for references. The NIR images (this section) and SED (Section~\ref{sec:sed}) are produced from simulations with $4 \times 10^7$ photon packets, and for the sub-mm images (Section~\ref{sec:submm}) we use $5 \times 10^8$ photon packets. By varying the random seeds in the Monte Carlo simulations, we find the noise levels in both the radial profile of the convolved images (Section~\ref{sec:psf}) and the SED to be $\lesssim0.5\%$ in the range of interest. In our models, we construct an axisymmetric disk (assumed to be at $\sim$140~pc) 200~AU in radius on a $600\times200$ grid in spherical coordinates ($R,\theta$), where $R$ is in the radial direction and $\theta$ is in the poloidal direction ($\theta=0^\circ$ is the disk mid-plane). We include accretion energy in the disk using the Shakura \& Sunyaev $\alpha$ disk prescription \citep{whi03a}. Disk accretion under the accretion rate assumed in our models below (several $\times10^{-9}M_\odot$ yr$^{-1}$) does not have a significant effect on the SED or the images (for simplicity, accretion energy from the inner gas disk is assumed to be emitted with the stellar spectrum, but see also the treatment in \citealt{ake05}). We model the entire disk with two components: a thick disk with small grains ($\sim$$\mu$m-sized and smaller, more of less pristine), and a thin disk with large (grown and settled) grains (up to $\sim$mm-sized). Figure~\ref{fig:sigma} shows the schematic surface density profile for both dust population.

The parametrized vertical density profiles for both dust populations are taken to be Gaussian (i.e.~$\rho(z)=\rho_0e^{-z^2/2h^2}$, isothermal in the vertical direction $z$), with scale heights $h_{\rm b}$ and $h_{\rm s}$ being simple power laws $h\propto R^\beta$ (we use subscripts ``s'' and ``b'' to indicate the small and big dust throughout the paper, while quantities without subscripts ``s'' and ``b'' are for both dust populations). Following A11, to qualitatively account for the possibility of settling of big grains, we fix $h_{\rm b}=0.2\times h_{\rm s}$ in most cases to simplify the models, unless indicated otherwise. Radially the disk is divided into two regions: an outer full disk from a cavity edge $R_{\rm cav}$ to 200~AU, and an inner cavity from the dust sublimation radius $R_{\rm sub}$ to $R_{\rm cav}$ ($R_{\rm sub}$ is determined self-consistently as where the temperature reaches the sublimation temperature $T_{\rm sub}\sim$1600~K, \citealt{dul01}, usually around 0.1$-$0.2~AU). At places in the disk where a large surface area of material is directly exposed to starlight, a thin layer of material is superheated, and the local disk ``puffs'' up vertically \citep{dul04a}. To study this effect at the inner rim ($R_{\rm sub}$) or at the cavity wall, we adopt a treatment similar to A11. In some models below we manually raise the scale height $h$ at $R_{\rm sub}$ or $R_{\rm cav}$ by a certain factor from its ``original'' value, and let the puffed up $h$ fall back to the underlying power law profile of $h$ within $\sim$0.1~AU as $e^{-(\delta R/0.1{\rm~AU})^2}$. We note that these puffed up walls are vertical, which may not be realistic \citep{ise05}.

For the surface density profile in the outer disk, we assume
\begin{equation}
\Sigma_{\rm o}(R)=\Sigma_{\rm cav}\frac{R_{\rm cav}}{R}e^{(R_{\rm cav}-R)/R_c}
\label{eq:sigmao}
\end{equation}
where $\Sigma_{\rm cav}$ is the surface density at the cavity edge (normalized by the total disk mass), $R_c$ is a characteristic scaling length, and the gas-to-dust ratio is fixed at 100. Following A11, we take $85\%$ of the dust mass to be in large grains at $R>R_{\rm cav}$. For the inner disk (i.e.~$R<R_{\rm cav}$) three surface density profiles have been explored:
\ba
&&\Sigma_{\rm i}(R)=(\delta_{\rm cav}\Sigma_{\rm cav})\frac{R_{\rm cav}}{R}e^{(R_{\rm cav}-R)/R_c}\ \ \ ({\rm rising}\ \Sigma_{\rm i}(R)),
\label{eq:sigmai-andrews} \\
&&\Sigma_{\rm i}(R)=\delta_{\rm cav}\Sigma_{\rm cav}\ \ \ ({\rm flat}\ \Sigma_{\rm i}(R)), {\rm\ and}
\label{eq:sigmai-flat} \\
&&\Sigma_{\rm i}(R)=(\delta_{\rm cav}\Sigma_{\rm cav}) \frac{R}{R_{\rm cav}}\ \ \ ({\rm declining}\ \Sigma_{\rm i}(R)),
\label{eq:sigmai-linearneg}
\ea
and their names are based on their behavior when moving inward inside the cavity. We note that Equation~\eqref{eq:sigmai-andrews} and \eqref{eq:sigmao} together form a single $\Sigma(R)$ scaling relation for the entire disk (with different normalization for the inner and outer parts), as in A11.

We define the depletion factor of the total dust inside the cavity as
\begin{equation}
\delta(R)=\frac{\Sigma_{\rm i}(R)}{\Sigma^{\rm full}_{\rm i}(R)},
\label{eq:deltacav}
\end{equation}
where $\Sigma^{\rm full}_{\rm i}(R)$ is found by extrapolating $\Sigma_{\rm o}(R)$ from the outer disk, i.e.~evaluating Equation~\eqref{eq:sigmao} at $R<R_{\rm cav}$ (or Equation~\eqref{eq:sigmai-andrews} with $\delta_{\rm cav}=1$). In addition, we define $\delta_{\rm s}(R)$ and $\delta_{\rm b}(R)$ as the cavity depletion factors for the small and big dust respectively as
\begin{equation}
\delta_{\rm s}(R)=\frac{\Sigma_{\rm i,s}(R)}{0.15\times\Sigma^{\rm full}_{\rm i}(R)},\ \delta_{\rm b}(R)=\frac{\Sigma_{\rm i,b}(R)}{0.85\times\Sigma^{\rm full}_{\rm i}(R)},
\label{eq:deltacav}
\end{equation}
where 0.15 and 0.85 are the mass fractions of the small and big dust in the outer disk. We note that unlike previous models such as A11, our cavity depletion factors are radius dependent (a constant $\delta_{\rm s}(R)$ or $\delta_{\rm b}(R)$ means a uniform depletion at all radii inside the cavity). Specifically, we define the depletion factor right inside the cavity edge as
\begin{equation}
\delta_{\rm cav}=\delta(R=R_{\rm cav}-\epsilon),\ \delta_{\rm cav,s}=\delta_{\rm s}(R=R_{\rm cav}-\epsilon),\ \delta_{\rm cav,b}=\delta_{\rm b}(R=R_{\rm cav}-\epsilon).
\label{eq:deltacav0}
\end{equation}
With the same $\delta_{\rm cav}$, different models with different $\Sigma_{\rm i}(R)$ profiles (Equations~\eqref{eq:sigmai-andrews}-\eqref{eq:sigmai-linearneg}) have similar $\Sigma_{\rm i}(R)$ (and $\delta(R)$) in the outer part of the cavity, but very different $\Sigma_{\rm i}(R)$ (and $\delta(R)$) at the innermost part. Lastly, the mass averaged cavity depletion factor $\langle\delta\rangle$ is defined as
\begin{equation}
\langle\delta\rangle=\frac{\int_0^{R_{\rm cav}}{\Sigma_{\rm i}(R)2\pi RdR}}{\int_0^{R_{\rm cav}}{\Sigma^{\rm full}_{\rm i}(R)2\pi RdR}}.
\label{eq:deltaint}
\end{equation}

SMA observations have placed strong constraints on the spatial distribution of the big dust, while the constraints on the small grains from the SED are less certain, especially beyond $R\sim10$~AU. Based on this, we adopt the spatial distribution of big grains in A11 (i.e.~Equation~\eqref{eq:sigmao}, and no big grains inside the cavity), and focus on the effect of the distribution of small grains inside the cavity. Therefore, the sub-mm properties of our models are similar to those of the models in A11 (Section~\ref{sec:submm}), since large grains dominate the sub-mm emission. We tested models with non-zero depletion for the big grains, and found that they make no significant difference as long as their surface density is below the SMA upper limit. From now on we drop the explicit radius dependence indicator $(R)$ from various quantities in most cases for simplicity.

\subsection{Dust properties}\label{sec:dust}

\begin{figure}[tb]
%\vspace*{-0.5cm}
\begin{center}
\epsscale{0.35} \plotone{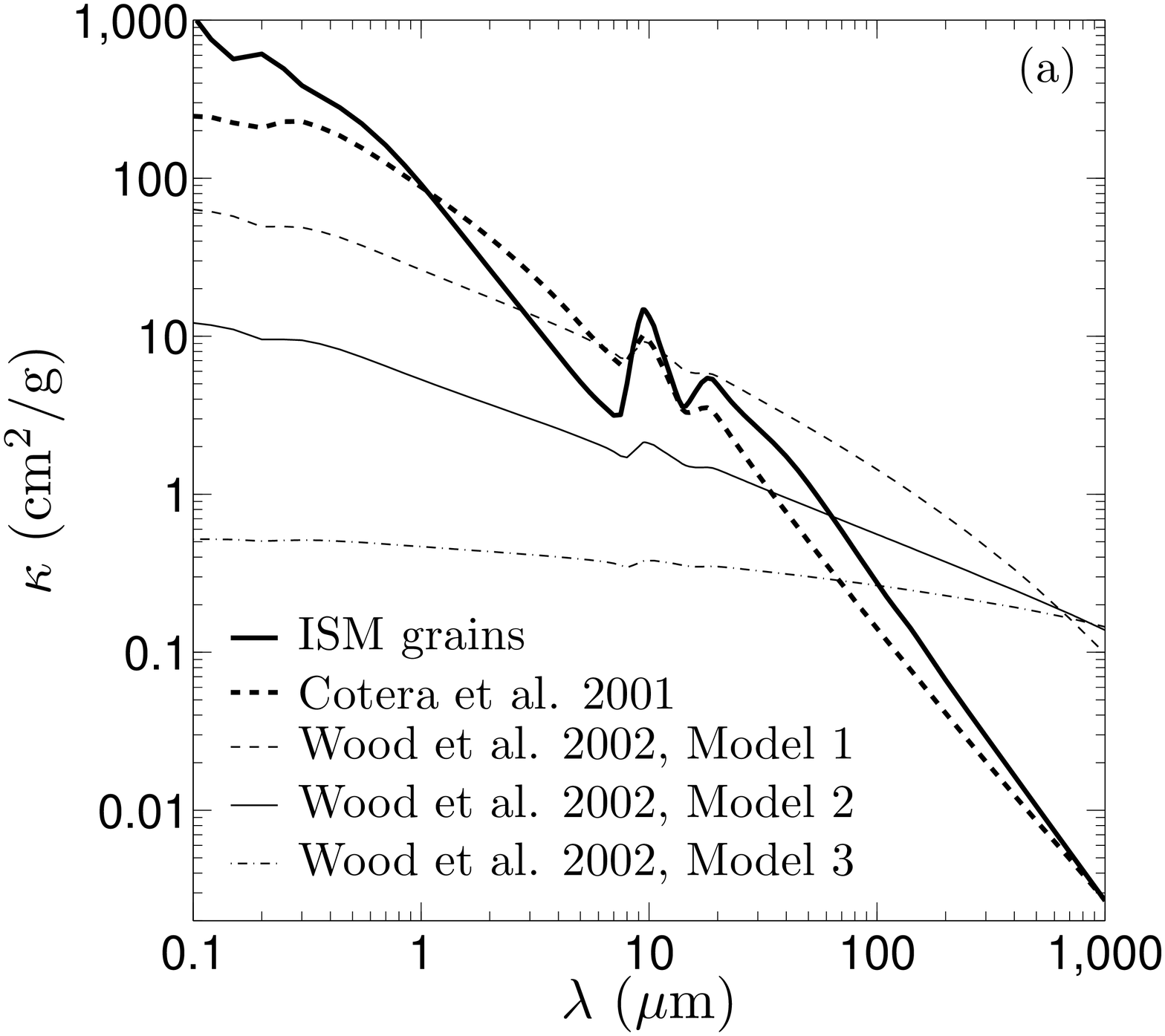} \plotone{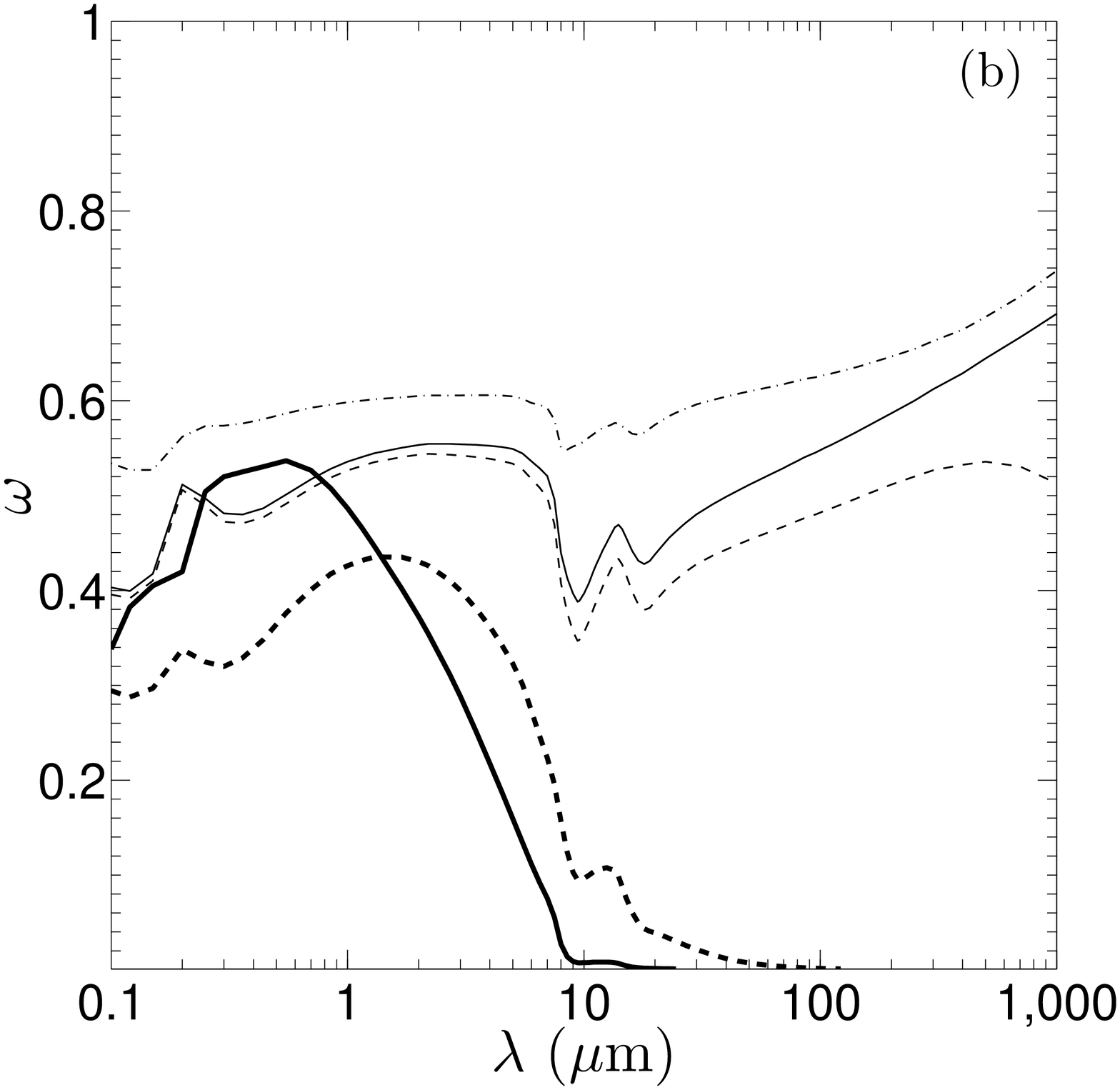} \plotone{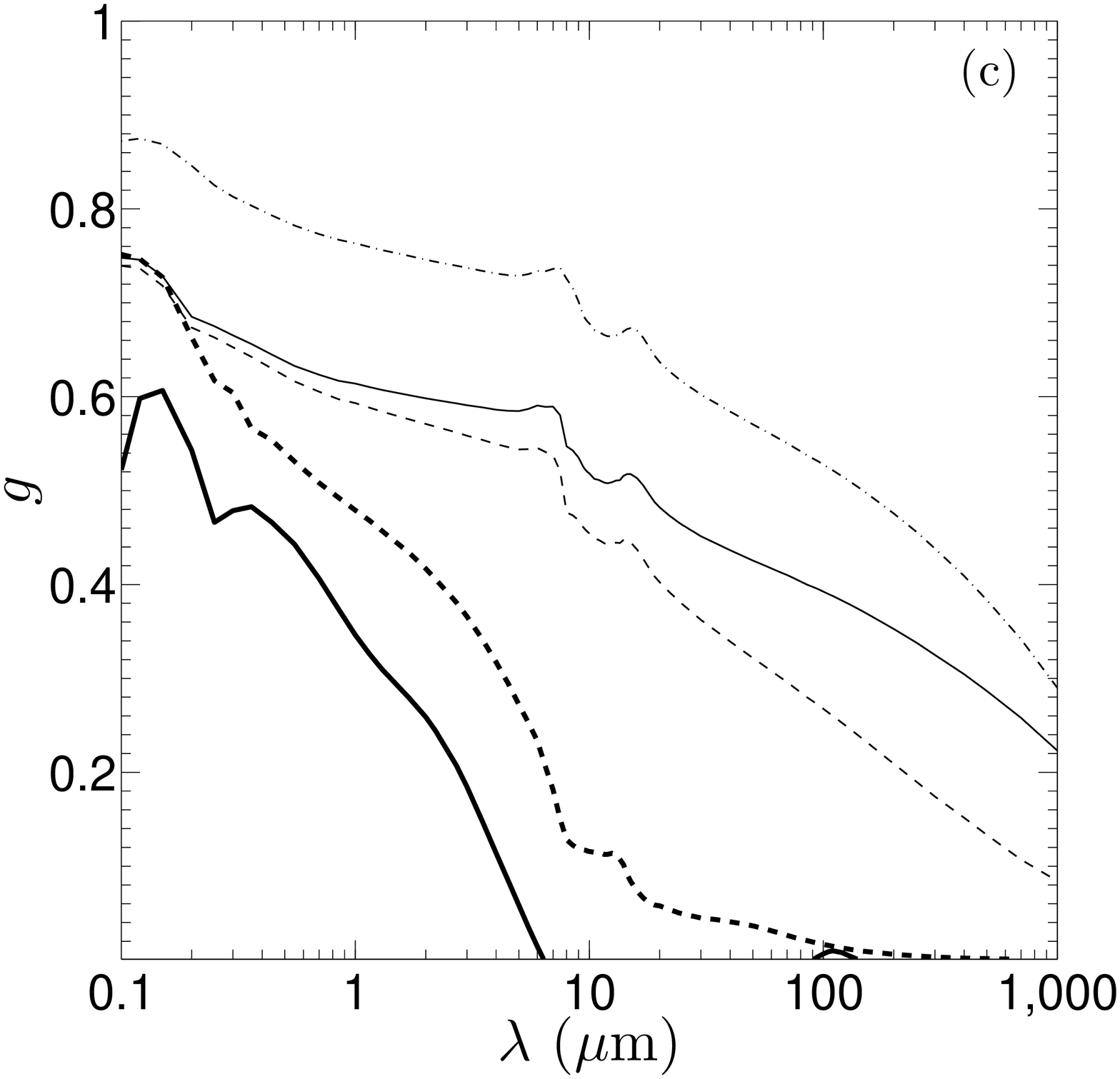} \plotone{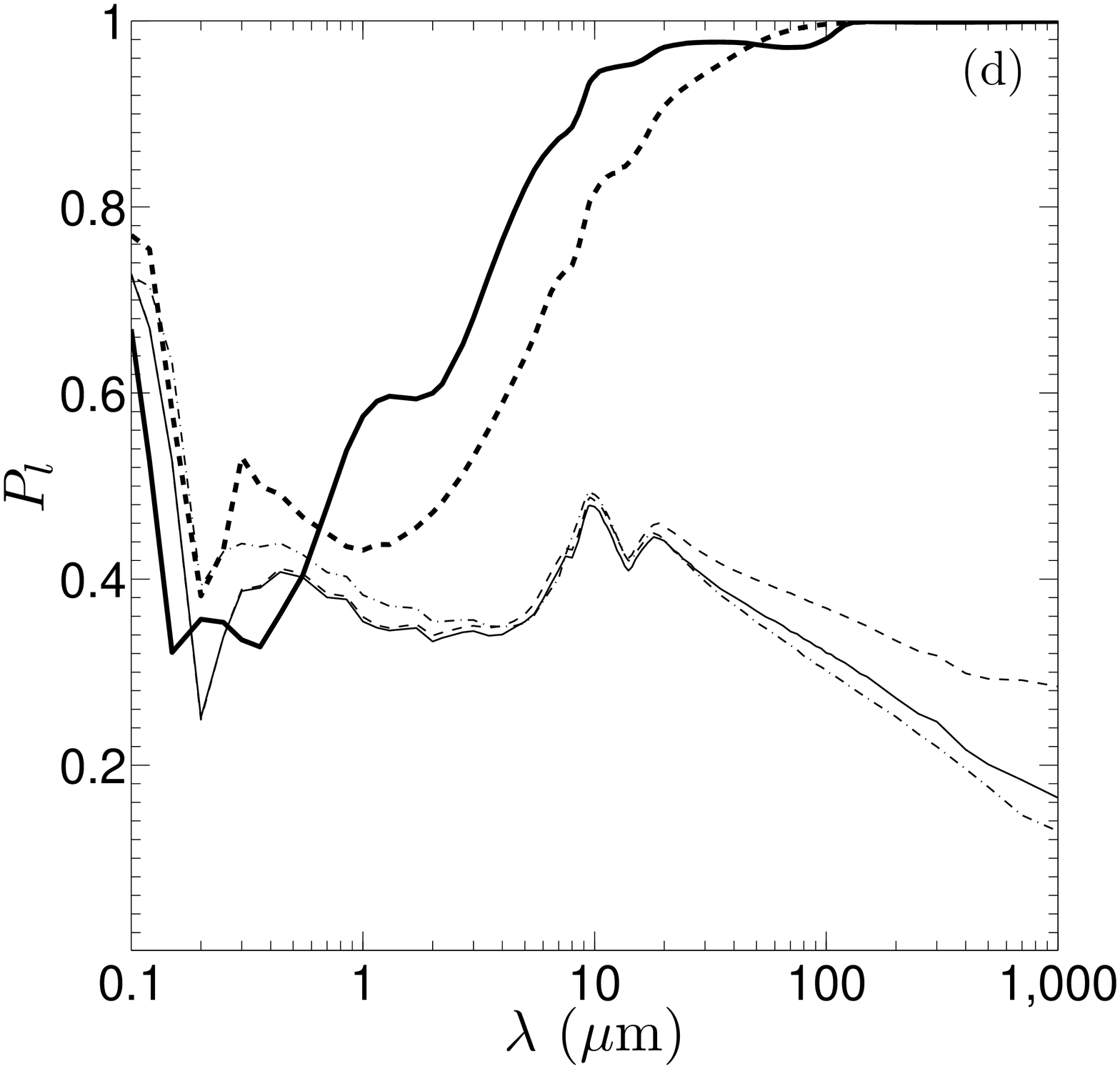}
%\vspace*{-0.5cm}
\end{center} \figcaption{Dust properties of the grain models we use, showing opacity $\kappa$ (a), albedo $\omega$ (b), average cosine scattering angle $g$ (c), and maximum polarization $P_l$ (d) (which typically occurs at a $90^\circ$ scattering angle). The opacity is the dust+gas opacity assuming a gas-to-dust ratio of 100. Thick curves are for the small dust models, which dominate the scattered light due to their large scale height and NIR opacity. Thin curves are for the big dust models, which dominate the sub-mm emission due to their high opacity at these wavelengths. The solid curves are the fiducial choices in this work.
\label{fig:dust}}
\end{figure}

For the small grains we try two models: the standard interstellar medium (ISM) grains (\citealt{kim94}, $\sim$micron-sized and smaller), and the model that \citet{cot01} employed to reproduce the HH 30 NIR scattered light images, which are somewhat larger than the ISM grains (maxim size $\sim$20~$\mu$m). These grains contain silicate, graphite, and amorphous carbon, and their properties are plotted in Figure~\ref{fig:dust}. The two grain models are similar to each other, and both are similar to the small grains model which A11 used in the outer disk and the cavity grains which A11 used inside the cavity and on the cavity wall. We note that for detailed modeling which aims at fitting specific objects, the model for the small grains needs to be turned for each individual object. For example, the strength and shape of the silicate features indicate different conditions for the small grains in the inner disk (\citealt{ada11}, and \citealt{fur11}, who also pointed out that the silicate features in transitional disks typically show that the grains in the inner disk are dominated by small amorphous silicate grains similar to ISM grains). However, since we do not aim at fitting specific objects, we avoid tuning the small dust properties and assume ISM grains \citep{kim94} for the models shown below, to keep our models generalized and simple.

For the large grains we try three different models, namely Models 1, 2, and 3 from \citet{woo02}. The properties of these models are plotted in Figure~\ref{fig:dust}. They adopt a power-law size distribution (i.e. as in \citealt{kim94}) with an exponential cutoff at large size, and the maxim size is $\sim$1~mm. These grains are made of amorphous carbon and astronomical silicates, with solar abundances of carbon and silicon. These models cover a large parameter space, however we find that they hardly make any difference in the scattered light image and the IRS SED, due to their small scale height and their absence inside the cavity. For this reason we fix our big grains as described by Model 2 in \citet{woo02} (which is similar to the model of the big grains in A11). We note that small grains have much larger opacity than big grains at NIR, and it is the other way around at sub-mm (Figure~\ref{fig:dust}).

\subsection{Post processing of the scattered light images}\label{sec:psf}

To obtain realistic images which can be directly compared to SEEDS observations, the raw NIR images of the entire disk+star system from the radiative transfer simulations need to be convolved with the point spread function (PSF) of the instrument. SEEDS can obtain both the FI and the PI images for any object, either with or without a coronagraph mask. The observation could be conducted in several different observational modes, including angular differential imaging (ADI, \citealt{mar06}), polarization differential imaging (PDI, \citealt{hin09}), and spectral differential imaging (SDI, \citealt{mar00}). For a description of the instrument see \citet{tam09} and \citet{suz10}.

In this work, we produce both the narrow band 880 $\mu$m images and $H$ band NIR images. While at 880 $\mu$m we produce the full intensity images, for the NIR scattered light images we focus on the PI images (produced in the PDI mode, both with and without a coronagraph mask). This is because (1) PDI is the dominate mode for this sample in SEEDS, and (2) it is more difficult to interpret FI (ADI) images since its reduction process partially or completely subtracts azimuthally symmetric structure. Other authors had to synthesize and reduce model data in order to test for the existence of features like cavities \citep{tha10} or spatially extended emission \citep{tha11}. For examples of PDI data reduction and analysis, see \citet{has11}. When observing with a mask, $\psi_{\rm in}$ in the PI images is the mask size (typically $0.\!\!''15$ in radius), and when observing without a mask, $\psi_{\rm in}$ is determined by the saturation radius, which typically is $\sim$$0.\!\!''1$.

To produce an image corresponding to observations made without a mask, we convolve the raw PI image of the entire system with an observed unsaturated HiCIAO $H$-band PSF. The resolution of the PSF is $\sim$$0.\!\!''05$ ($\sim$1.2$\lambda/D$ for an 8-m telescope) and the Strehl ratio is $\sim$40\% \citep{suz10}. The integrated flux within a circle of radius $0.\!\!''25$ is $\sim$80\% of the total flux ($\sim$90\% for a circle of radius $0.\!\!''5$). We then carve out a circle at the center with $0.\!\!''1$ in radius to mimic the effect of saturation. We call this product the convolved unmasked PI image. To produce an image corresponding to observations with a mask, we first convolve the part of the raw PI image which is not blocked by the mask with the above PSF. We then convolve the central source by an observed PI coronagraph stellar residual map (the PSF under the coronagraph), and add this stellar residual to the disk images (the flux from the inner part of the disk which is blocked by the mask, $\sim$20~AU at $\sim$140 pc, is added to the star). Lastly, we carve out a circle $0.\!\!''15$ in radius from the center from the combined image to indicate the mask. We call this product the convolved masked PI image. We note that the stellar residual is needed to fully reproduce the observations, but in our sample the surface brightness of the stellar residual is generally well below the surface brightness of the disk at the radius of interest, so it doesn't affect the properties of the images much.

In this study, the disk is assumed to be face-on in order to minimize the effect of the phase function in the scattering, so that we can focus on the effect of the disk structure. This is a good approximation since most objects in this IRS/SMA/Subaru sample have inclinations around $\sim$25$^\circ$ (i.e.~minor to major axis ratio $\sim$0.9. An observational bias towards face-on objects may exist, since they are better at revealing the cavity). Additional information about the scattering properties of the dust could be gained from analyzing the detailed azimuthal profile of the scattered light in each individual system, which we defer to the future studies. To calculate the azimuthally averaged surface brightness profiles, we bin the convolved images into a series of annuli $0.\!\!''05$ in width (the typical spatial resolution), and measure the mean flux within each annulus.

%%%%%%%%%%%%%%%%%%%%%%%%%%%%%%%%%%%%%%%%%%%%%%%%%%%%

%%%%%%%%%%%%%%%%%%%%%%%%%%%%%%%%%%%%%%%%%%%%%%%%%%%%

\section{The NIR polarized scattered light images}\label{sec:image}

With the tools described above, we investigate what kinds of disk structure could simultaneously reproduce the gross properties of all three kinds of observations described in Section~\ref{sec:introduction}. In this section, we first investigate the properties of the scattered light images from a semi-analytical theoretical point of view (Section~\ref{sec:imagetheory}), then we present the model results from the Monte Carlo simulations (Section~\ref{sec:imageresult}).

\subsection{Theoretical considerations}\label{sec:imagetheory}

In a single NIR band, when the (inner) disk is optically thick (i.e. not heavily depleted of the small grains), the scattering of the starlight can be approximated as happening on a scattering surface $z_{\rm s}$ where the optical depth between the star and surface is unity (the single scattering approximation). This surface is determined by both the disk scale height (particularly $\beta$ in the simple vertically-isothermal models), and the radial profile of the surface density of the grains. The surface brightness of the scattered light $I_{\rm scat}(R)$ scales with radius as \citep{jan03,ino08}
\begin{equation}
I_{\rm scat}(R)\sim \zeta p\frac{L_\star}{4\pi R^2}\sin{\gamma}
\label{eq:ir}
\end{equation}
where $L_\star$ is the stellar luminosity at this wavelength, $\zeta$ is a geometrical scattering factor, $p$ is the polarization coefficient for PI ($p=1$ for FI), and $\gamma$ is the grazing angle (the angle between the impinging stellar radiation and the tangent of the scattering surface). We note that both $\zeta$ and $p$ depend on azimuthal angle, inclination of the disk, and the scattering properties of the specific dust population responsible for scattering at the particular wavelength. However, if the disk is relatively face-on and not too flared, they are nearly position independent, because the scattering angle is nearly a constant throughout the disk and the dust properties of the specific dust population do not change much with radius.

The grazing angle is determined by the curvature of the scattering surface. In axisymmetric disks, assuming the surface density and scale height of the small dust to be smooth functions of radius, the grazing angle is also smooth with radius, and two extreme conditions can be constrained as follows:
\begin{enumerate}
\item For disks whose scattering surface is defined by a constant poloidal angle $\theta$ (such as a constant opening angle disk), $\gamma\sim R_\star/R$ (\citealt{chi97}, where $R_\star$ is the radius of the star), so the brightness of the scattered light scales with $R$ as
\begin{equation}
I_{\rm scat}(R)\propto R^{-3}.
\label{eq:ir-3}
\end{equation}
\item For flared disks (but not too flared, $h\lesssim R$), the grazing angle can be well approximated as
\begin{equation}
\sin{\gamma}\sim\gamma\sim \frac{dz_{\rm s}}{dR}-\frac{z_{\rm s}}{R}.
\label{eq:gamma}
\end{equation}
In this case, \citet{mut11} explicitly calculated the position of the scattering surface $z_{\rm s}$ at various radii (see also \citealt{chi97}), and found $z_{\rm s}=\eta h_{\rm s}$ where the coefficient $\eta\sim$ a few and is nearly a constant. Combined with the fact that $h_{\rm s}/R\propto R^{\beta-1}$ ($\beta\sim1.25-1.3$ as typical values in irradiated disks, \citealt{chi97,har98}), we have the intensity of the scattered light scales with radius as
\begin{equation}
I_{\rm scat}(R)\propto R^{\beta-3}.
\label{eq:ir-2}
\end{equation}
\end{enumerate}

Although the above calculations are under two extreme conditions (for complete flat or flared disks), and they are based on certain assumptions and the observed images have been smeared out by the instrument PSF, the radial profiles (azimuthally averaged, or along the major axis in inclined systems) of SEEDS scattered light images for many objects in this sample lie between Equations (\ref{eq:ir-3}) and (\ref{eq:ir-2}) in the radius range of interest. In order to guide the eye and ease the comparison between modeling results and observations, we use the scaling relation
\begin{equation}
I_{\rm scat}(R)\propto R^{-2.5}
\label{eq:seeds}
\end{equation}
to represent typical observational results, and plot it on top of the radiative transfer results, which will be presented in Section~\ref{sec:imageresult} (with arbitrary normalization).

On the other hand, an abrupt jump in the surface density or scale height profile of the small dust in the disk produces a jump in $\gamma$ at the corresponding position. The effect of this jump will be explored in Section~\ref{sec:imageresult}.

\subsection{Modeling results}\label{sec:imageresult}

\begin{figure}[tb]
%\vspace*{-0.5cm}
\begin{center}
\epsscale{0.6} \plotone{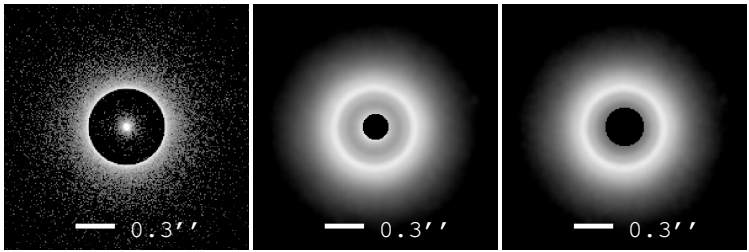}
%\vspace*{-0.5cm}
\end{center} \figcaption{Simulated $H$ band scattered light PI images for a face-on disk with a uniformly heavily depleted ($\delta_{\rm s}\approx10^{-5}$) cavity (see Section~\ref{sec:imageresult} for details), displayed with logarithmic stretches. The left panel shows the raw disk image from Whitney's radiative transfer simulation, the middle panel shows the unmasked image convolved with the HiCIAO PSF, and the right panel shows the convolved masked image (see Section~\ref{sec:psf} for details on the HiCIAO PSF). The cavity is 42~AU in radius ($\sim$$0.\!\!''3$ at 140 pc), and the inner working angles are $\sim$$0.\!\!''1$ without a coronagraph, and $\sim$$0.\!\!''15$ with a coronagraph (the hole in the middle and right panels, respectively). The central surface brightness deficit and the bright ring at the cavity edge are prominent under both observation modes, which is qualitatively inconsistent with many SEEDS transitional disks observations. This indicates that the cavity is not heavily depleted in small dust.
\label{fig:image-andrews}}
\end{figure}

\begin{figure}[tb]
%\vspace*{-0.5cm}
\begin{center}
\epsscale{0.4} \plotone{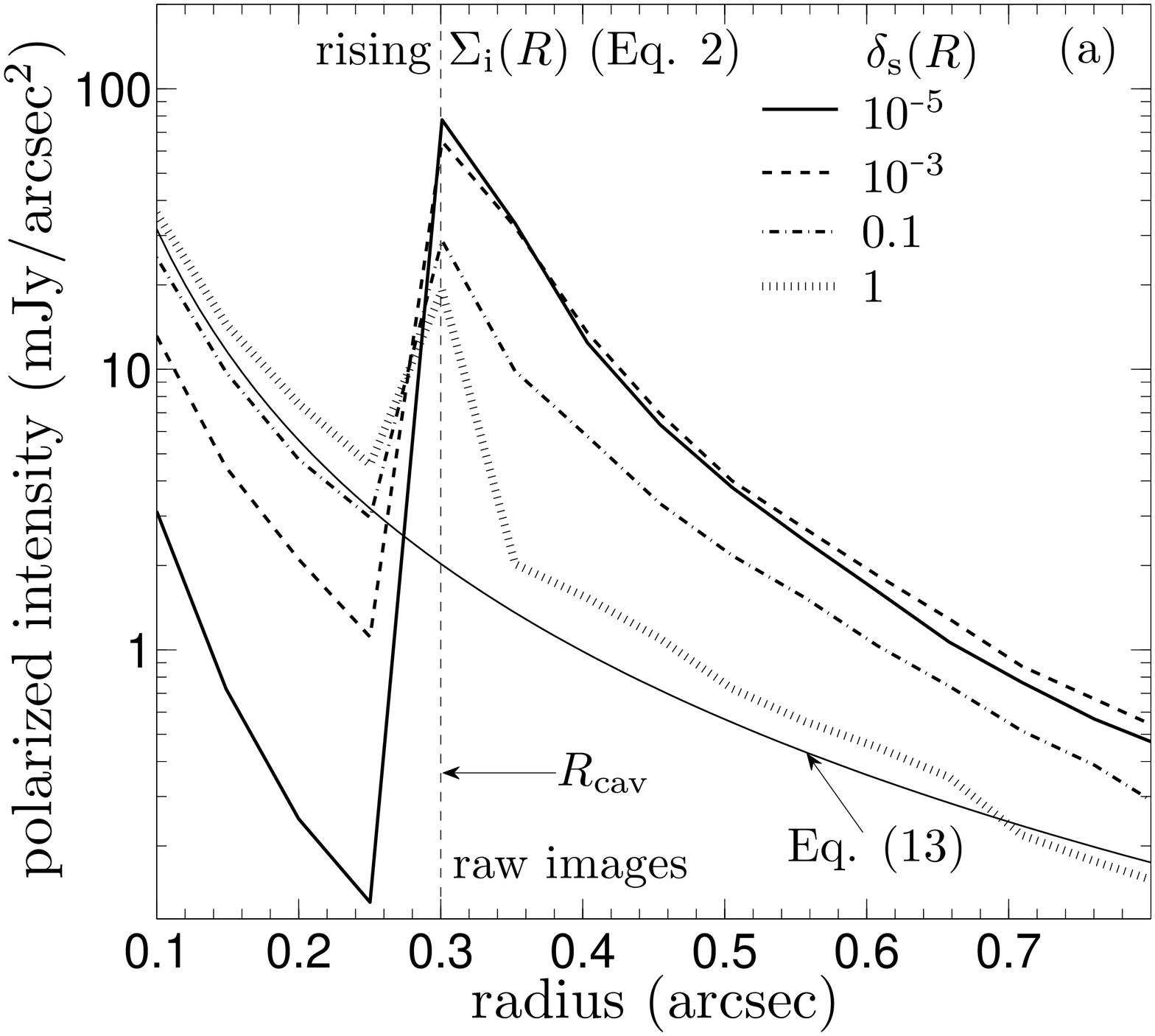} \plotone{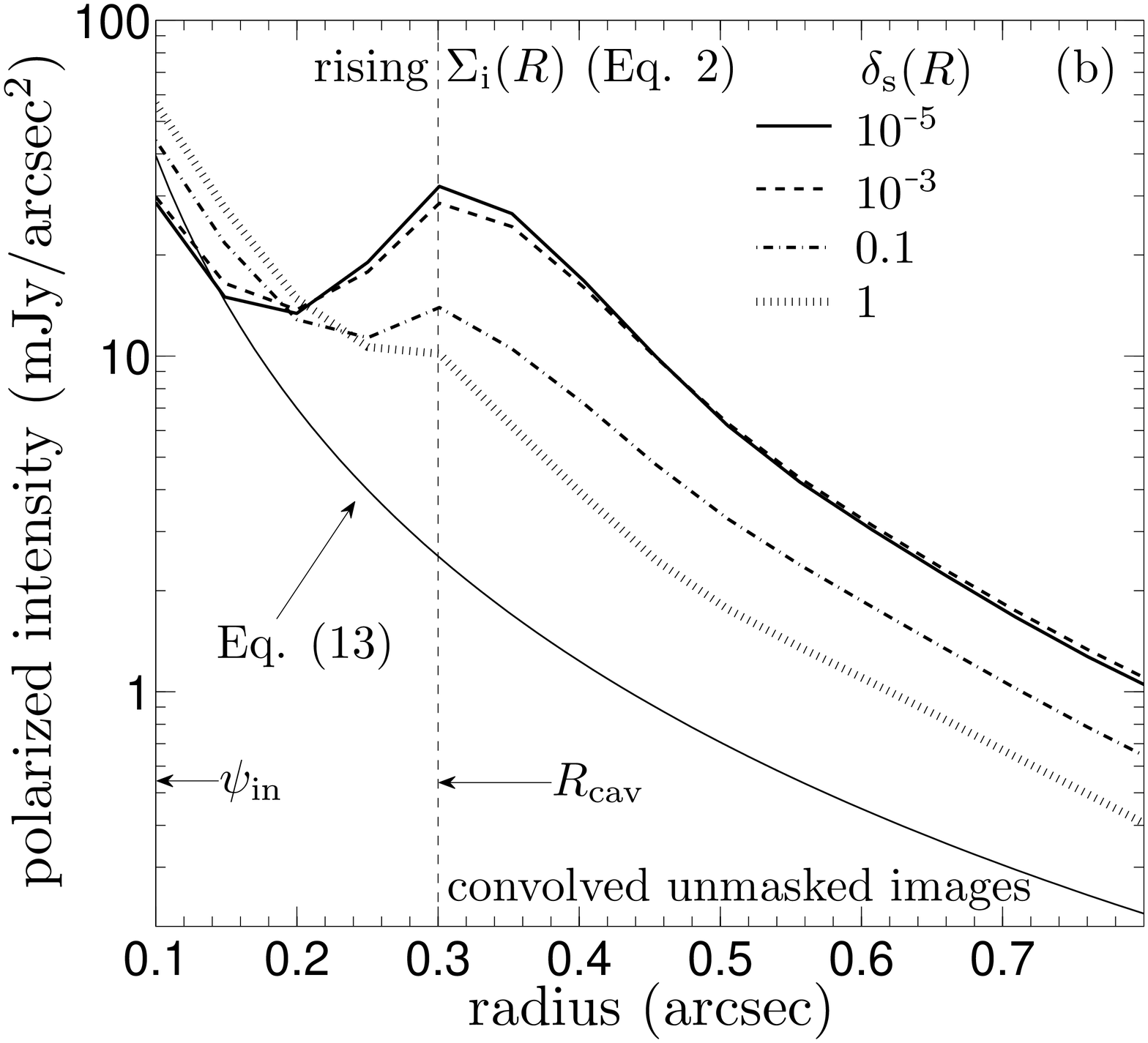} \plotone{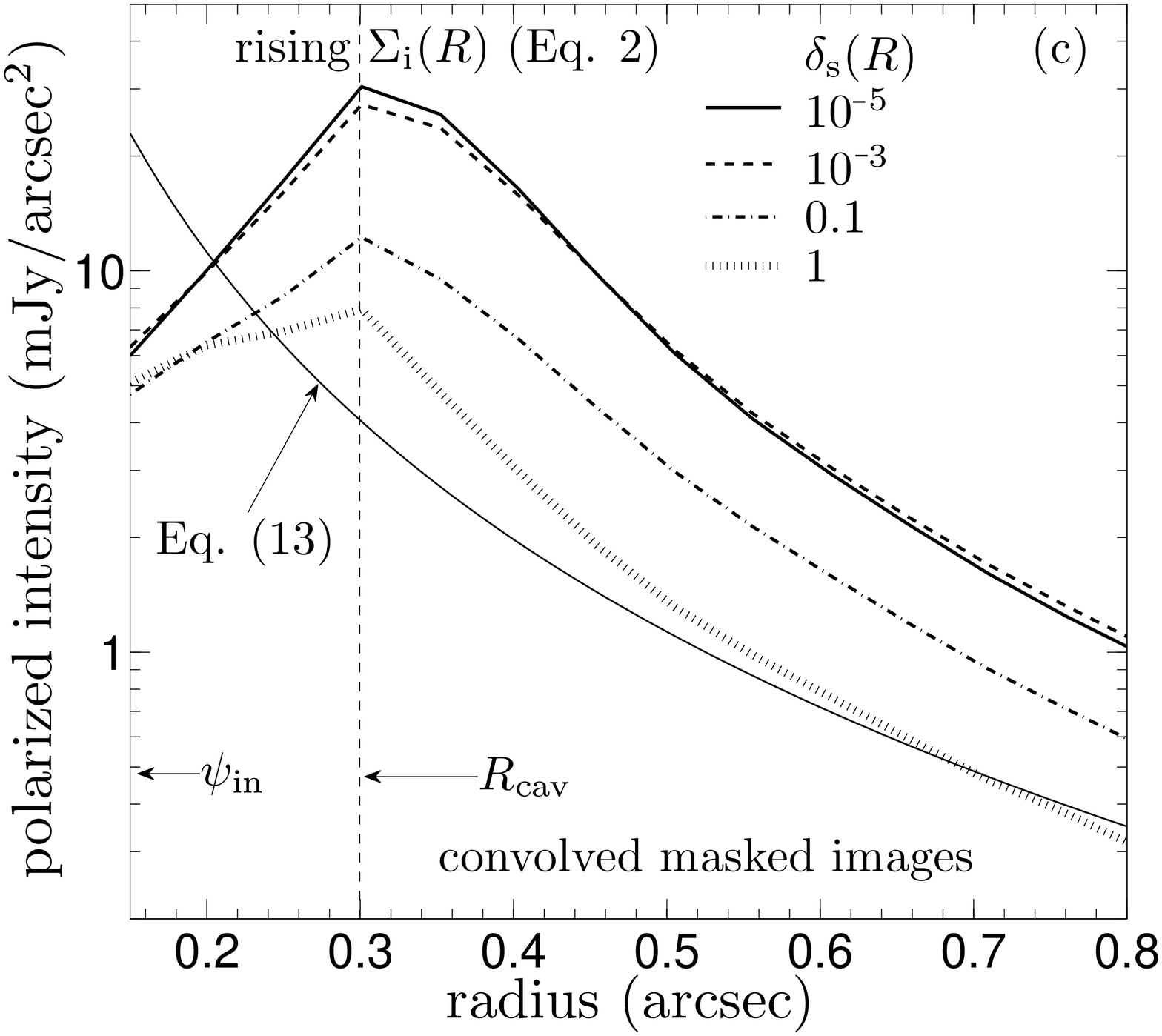}
%\vspace*{-0.5cm}
\end{center} \figcaption{Surface brightness radial profiles of the raw images (a), convolved unmasked PI images (b), and convolved masked PI images (c), assuming a cavity completely devoid of large dust and uniformly depleted in small dust. The model with $\delta_{\rm s}=10^{-5}$ (thick solid curves) corresponds to Figure~\ref{fig:image-andrews} and typical models in A11, while the $\delta_{\rm s}=1$ model (dotted curves) has no depletion in small grains at the cavity edge. The thin solid curves labeled as Equation~\eqref{eq:seeds} represent the scaling relation typical in SEEDS (with arbitrary normalization). The basic features of the uniformly heavily depleted models --- an increase in intensity at the cavity edge and a drop inside --- are inconsistent with typical SEEDS results, which smoothly increase in intensity up to the inner working angle.
\label{fig:rp-andrews}}
\end{figure}

First, we present the simulated $H$ band PI images for a face-on transitional disk with a uniformly heavily depleted cavity with $\Sigma_{\rm i}$ as Equation~\eqref{eq:sigmai-andrews} (Figure~\ref{fig:image-andrews}), and the associated surface brightness radial profiles (the thick solid curves in Figure~\ref{fig:rp-andrews}). In each figure, the three panels show the raw image, the convolved unmasked image, and the convolved masked image, respectively. This model is motivated by the disk+cavity models in A11; we therefore use parameters typical of those models. Experiments show that the peculiarities in each individual A11 disk model hardly affect the qualitative properties of the images and their radial profiles, as long as the cavity is large enough ($\gtrsim0.\!\!''2$) and the disk is relatively face-on.

This disk harbors a giant cavity at its center with $R_{\rm cav}=42$~AU ($\sim$$0.\!\!''3$ at 140 pc). The disk has the same inwardly rising $\Sigma$ scaling both inside and outside the cavity as Equations (\ref{eq:sigmao}) and (\ref{eq:sigmai-andrews}). Outside the cavity both dust populations exist, with the big/small ratio as $0.85/0.15$. Inside the cavity there is no big dust ($\delta_{\rm b}=0$), and the small dust is uniformly heavily depleted to $\delta_{\rm s}=\delta_{\rm cav,s}\approx10^{-5}$ ($\delta\approx10^{-6}$). The surface density profiles for both dust populations can be found in Figure~\ref{fig:sigma}. The disk has total mass of $0.01M_\odot$ (gas-to-dust mass ratio 100), $h/R=0.075$ at 100~AU with $\beta=1.15$, $R_c=40$~AU, and accretion rate $\dot{M}=5\times10^{-9}M_\odot$ yr$^{-1}$. The central source is a 3 $R_\odot$, $2M_\odot$, 5750~K G3 pre-main sequence star. We puff up the inner rim and the cavity wall by 100\% and 200\%, respectively.

The most prominent features of this model in both the unmasked and masked PI images are the bright ring at $R_{\rm cav}$, and the surface brightness deficit inside the ring (i.e. the cavity). Correspondingly, the surface brightness profile increases inwardly in the outer disk, peaks around $R_{\rm cav}$, and then decreases sharply. This is very different from many SEEDS results (such as the examples mentioned in Section~\ref{sec:introduction}), in which both the bright ring and the inner deficit are absent, and the surface brightness radial profile keeps increasing smoothly all the way from the outer disk to the inner working angle, as illustrated by the scaling relation (\ref{eq:seeds}) in Figure~\ref{fig:rp-andrews}.

This striking difference between models and observations suggests that the small dust cannot have such a large depletion at the cavity edge. Figure~\ref{fig:rp-andrews} shows the effect of uniformly filling the cavity with small dust on the radial profile for unmasked images (left) and masked images (right), leaving the other model parameters fixed. As $\delta_{\rm s}$ gradually increases from $\sim$$10^{-5}$ to 1, the deviation in the general shape between models and observations decreases. The $\delta_{\rm s}=1$ model corresponds to no depletion for small grains inside the cavity (i.e.~a full small dust disk); this model agrees much better with the scattered light observations, despite a bump around the cavity edge produced by its puffed up wall (Section~\ref{sec:image-puffingup}), although this model fails to reproduce the transitional-disk-like SED (Section~\ref{sec:sed}).

We note that this inconsistency between uniformly heavily depleted cavity models and observations is intrinsic and probably cannot be solved simply by assigning a high polarization to the dust inside the cavity. The polarization fraction (PI/FI) in the convolved disk images of our models ranges from $\sim$0.3 to $\sim$0.5, comparable to observations (for example \citealt{mut12}). Even if we artificially increase the polarization fraction inside the cavity by a factor of 10, by scaling up the cavity surface brightness in the raw PI images (which results in a ratio PI/FI greater than unity) while maintaining the outer disk unchanged, the convolved PI images produced by these uniformly heavily depleted cavity models still have a prominent cavity at their centers. We also note that the contrast of the cavity (the flux deficit inside $R_{\rm cav}$) and the strength of its edge (the brightness of the ring at $R_{\rm cav}$) are partially reduced in the convolved images compared with the raw images. This is due both to the convolution of the disk image with the telescope PSF, which naturally smooths out any sharp features in the raw images, and to the superimposed seeing halo from the bright innermost disk (especially for the unmasked images).

\begin{figure}[tb]
%\vspace*{-0.5cm}
\begin{center}
\epsscale{0.6} \plotone{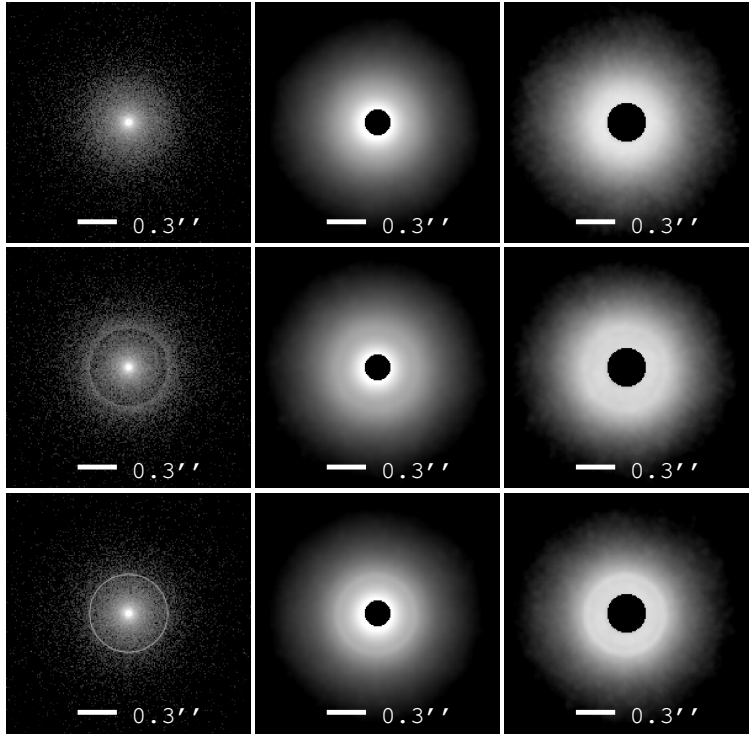}
%\vspace*{-0.5cm}
\end{center} \figcaption{Simulated $H$ band scattered light PI images for three face-on disks with a 42 AU cavity ($\sim$$0.\!\!''3$ at 140 pc), which is devoid of big dust and filled with a uniform surface density of small dust (flat $\Sigma_{\rm i}$, Equation~\eqref{eq:sigmai-flat}), displayed with logarithmic stretches. Panels (from left to right) are raw images, convolved unmasked images, and convolved masked images (the same as in Figure~\ref{fig:image-andrews}). The top and middle rows show cavities undepleted and 70\% depleted in small dust at the edge ($\delta_{\rm cav,s}=1$ and $\delta_{\rm cav,s}=0.3$, respectively), corresponding to the solid curves in Figure~\ref{fig:image-andrews}. The bottom row shows a cavity undepleted in small dust at the edge but with the cavity wall puffed up by 200\%. For detailed model parameters see Section~\ref{sec:imageresult}. Images in the top row are qualitatively similar to many SEEDS transitional disk observations in this sample, which are smooth on large scales and lack abrupt breaks or bumps at the cavity edge. Also, results show that both a modest surface density discontinuity (a $70\%$ drop, middle row) and a puffed up cavity wall (bottom row) could well be visible.
\label{fig:image-variation}}
\end{figure}

\begin{figure}[tb]
%\vspace*{-0.5cm}
\begin{center}
\epsscale{0.325} \plotone{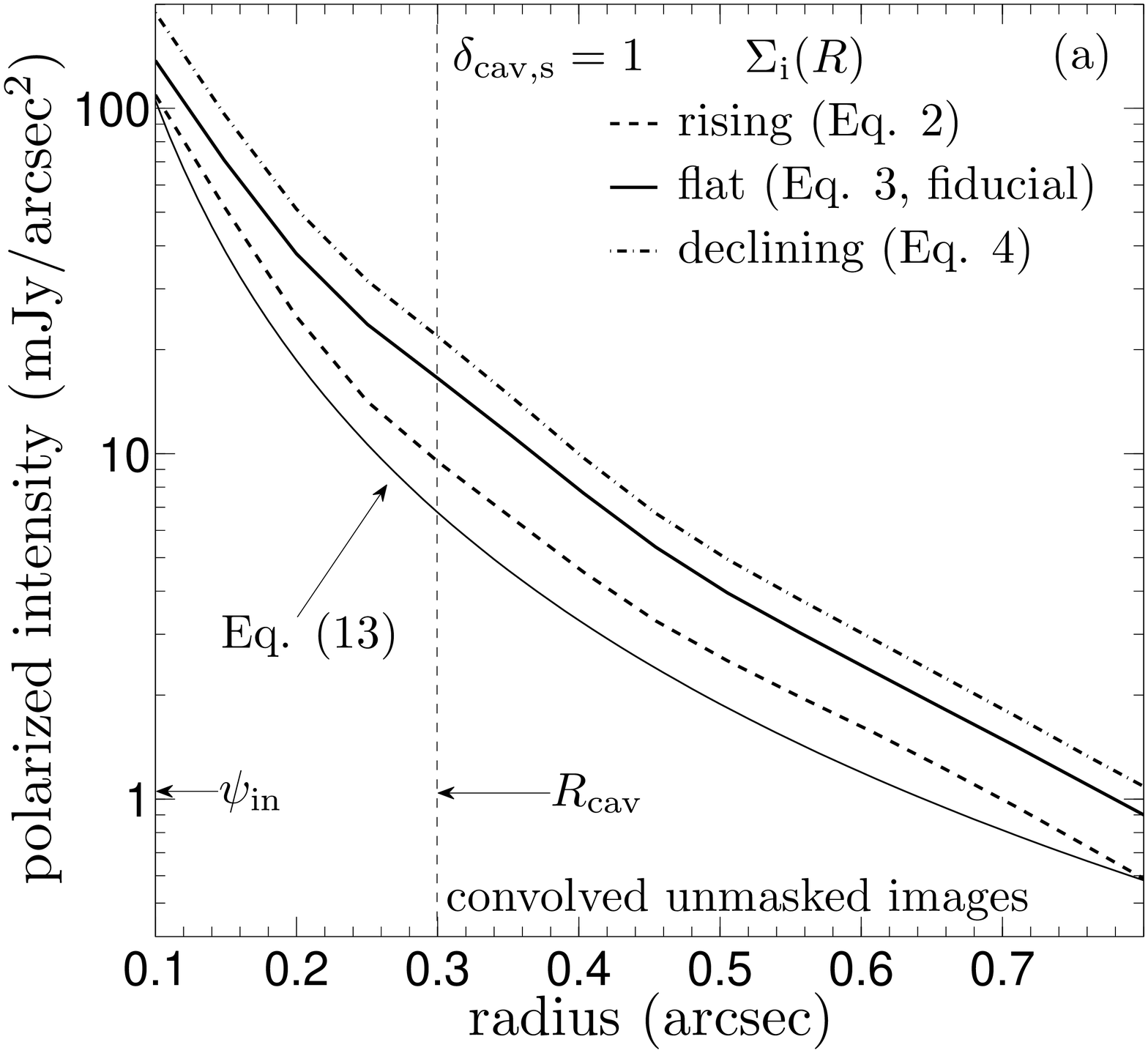} \plotone{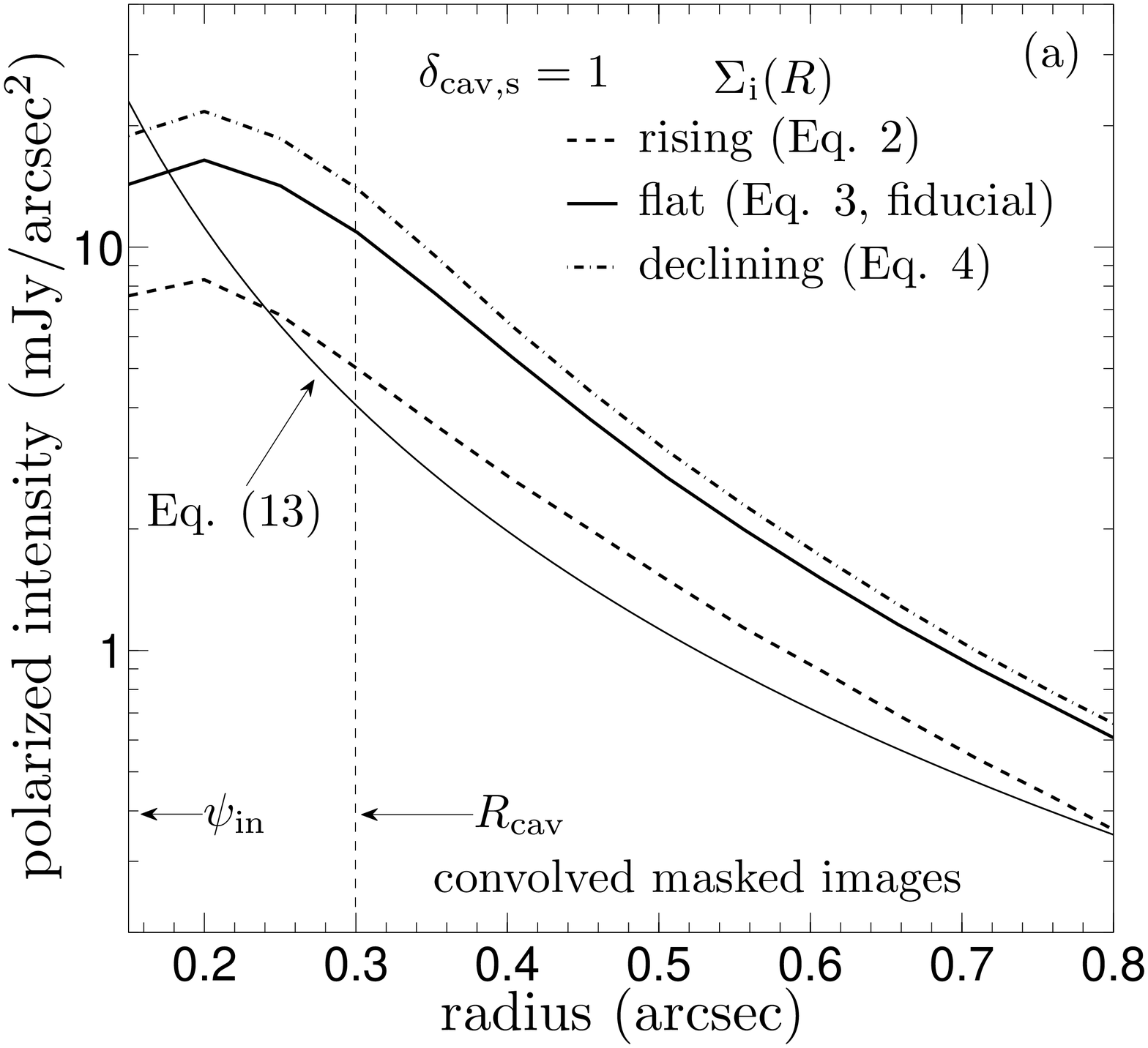} \plotone{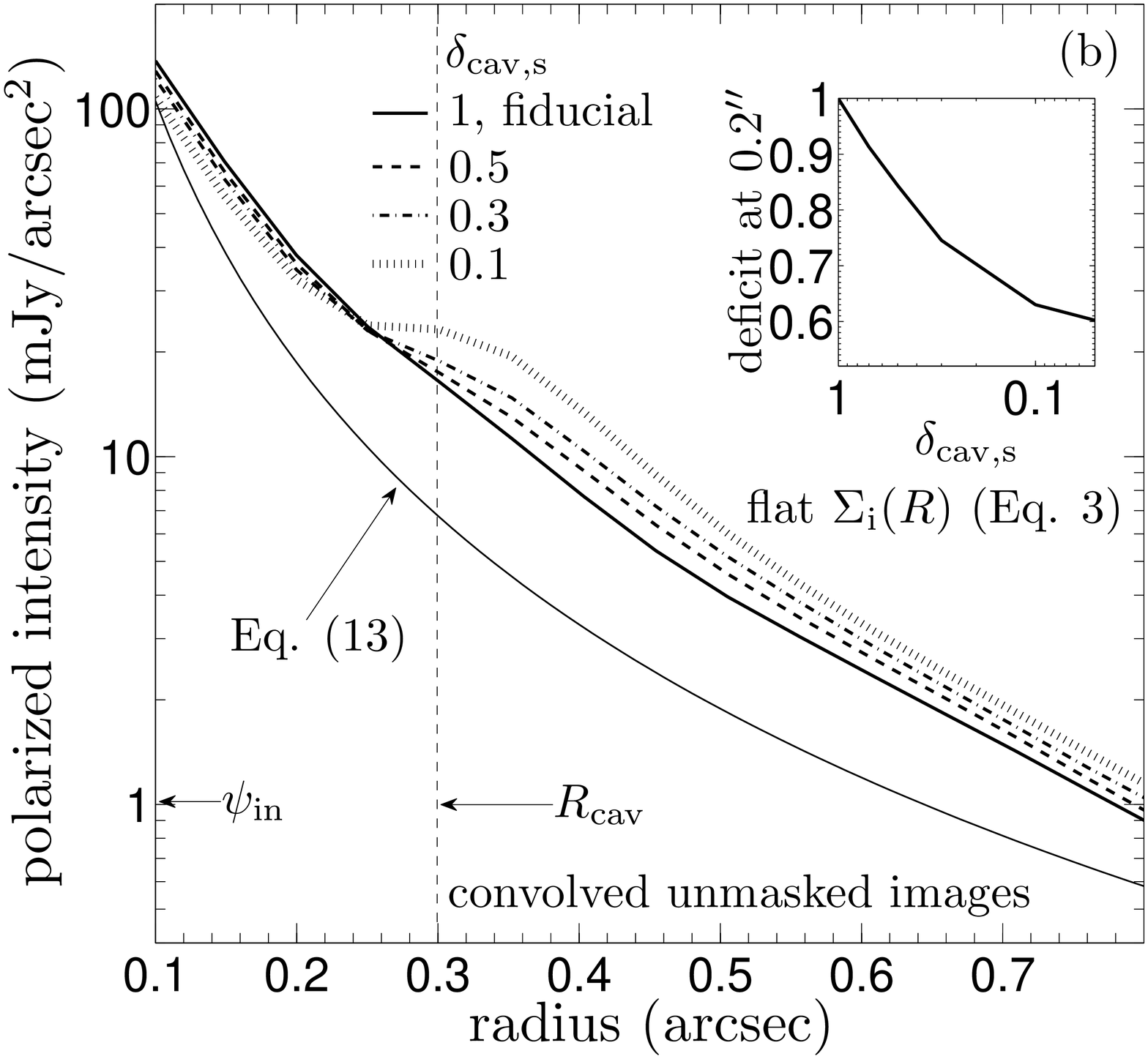} \plotone{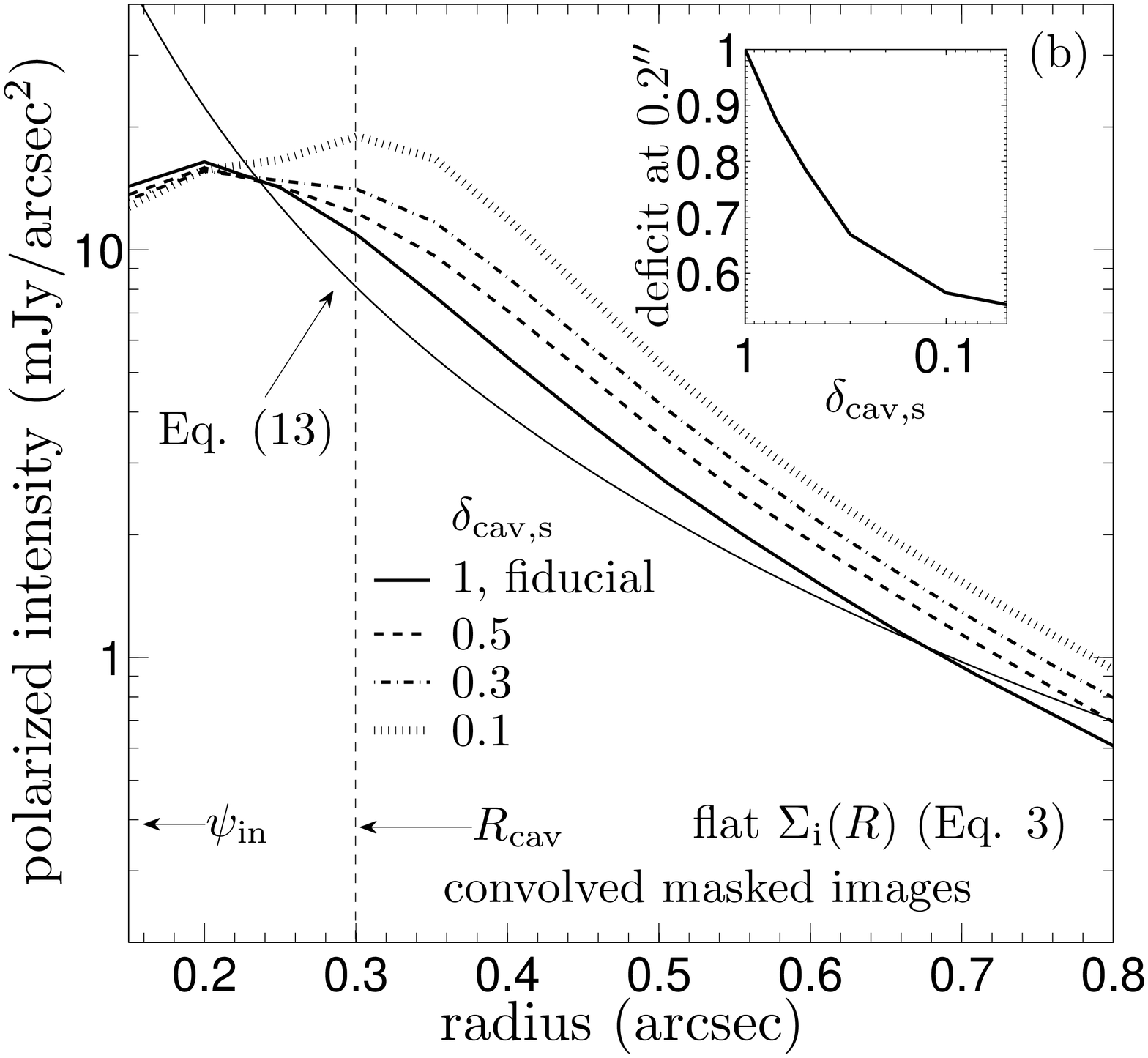} \plotone{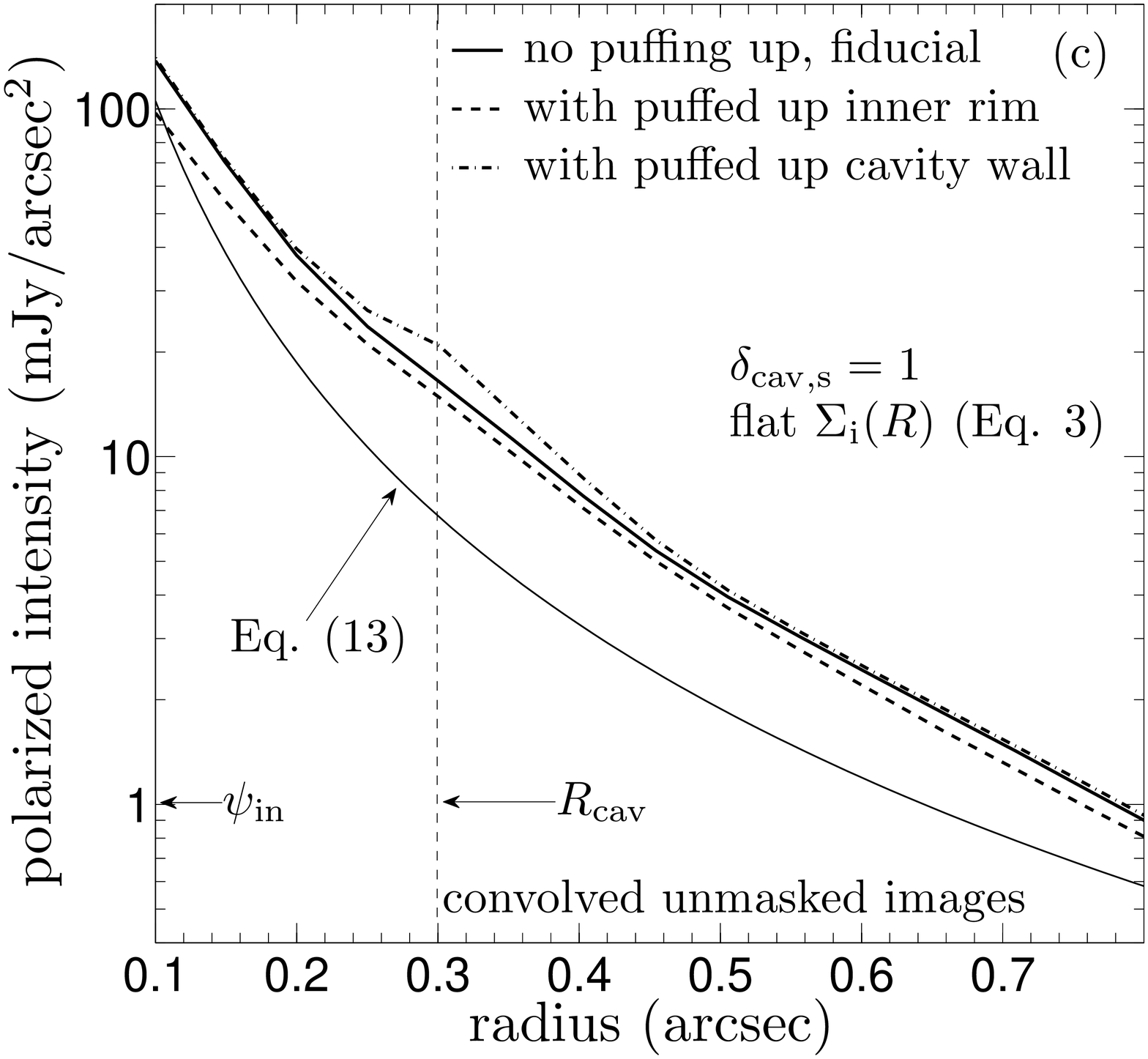} \plotone{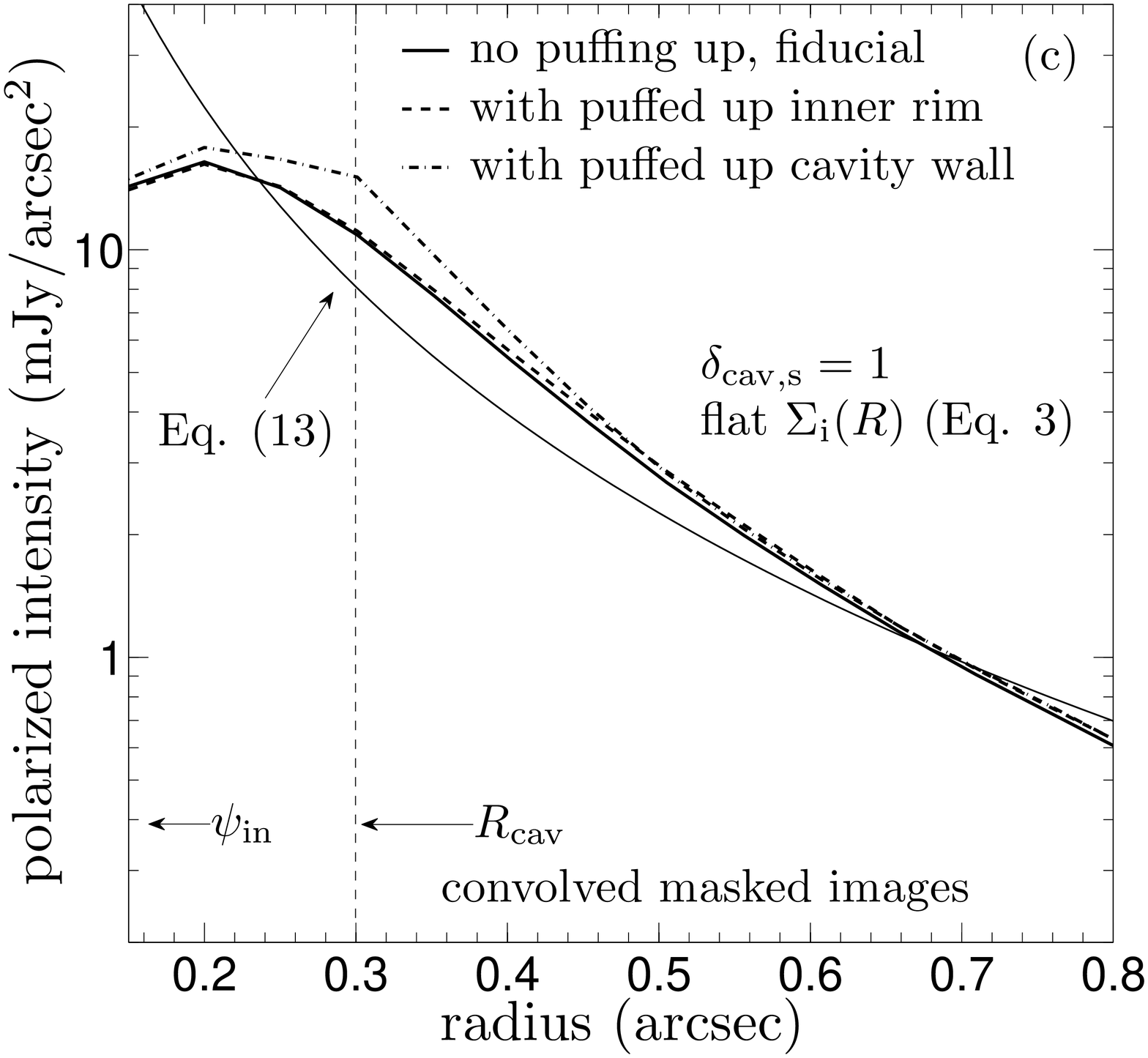}
%\vspace*{-0.5cm}
\end{center} \figcaption{Surface brightness radial profiles of the convolved unmasked (left) and masked (right) PI images for the models in Section~\ref{sec:imageresult}, showing the effect of $\Sigma_{\rm i}$ (a), $\delta_{\rm cav,s}$ (b), and the puffed up inner rim and cavity wall (c). The subpanel in panel (b) shows the relative flux deficit at $0.\!\!''2$ ($\frac{2}{3} R_{\rm cav}$) as a function of $\delta_{\rm cav,s}$ (Section~\ref{sec:image-deltacav015}). All models have no big dust inside the cavity. The curves labled ``fiducial'' are the same in all panels, which is from a model with flat $\Sigma_{\rm i}$, a continuous small dust surface density profile ($\delta_{\rm cav,s}=1$), and no puffed up cavity wall (the top row in Figure~\ref{fig:image-variation}). Results show that both a modest depletion in the small grains at $R_{\rm cav}$ and a puffed up cavity wall can add a bump to the originally smooth radial profile, while various forms of $\Sigma_{\rm i}$ with no discontinuity at $R_{\rm cav}$ produce similar surface brightness radial profile. Typical SEEDS results, as summarized in Section~\ref{sec:introduction} and illustrated by the scaling relation~\eqref{eq:seeds} (with arbitrary normalization), is consistent with smooth small dust disks continuous in both surface density and scale height. The apparent inward surface brightness decrease near $\psi_{\rm in}$ in the masked images is a coronagraph edge effect (Section~\ref{sec:image-sigmai}).
\label{fig:rp-variation}}
\end{figure}

In the rest of Section~\ref{sec:image} we focus on the small grains inside the cavity while keeping the big grains absent, and study the effect of the parameters $\Sigma_{\rm i}$, $\delta_{\rm cav,s}$, and the puffing up of the inner rim and cavity wall on the scattered light images. The convolved images for three representative models are shown in Figure~\ref{fig:image-variation}, and the radial profiles for all models are shown in Figure~\ref{fig:rp-variation}. Each model below is varied from one standard model, which is shown as the top panel in Figure~\ref{fig:image-variation} and represented by the thick solid curve in all panels in Figure~\ref{fig:rp-variation}. This fiducial model has a flat $\Sigma_{\rm i}$ with no discontinuity at the cavity edge for the small dust (i.e.~$\delta_{\rm cav,s}=1$), and no puffed up rim or wall, but otherwise identical parameters to the models above.

\subsubsection{The effect of $\Sigma_{\rm i}$}\label{sec:image-sigmai}

First we study the effect of three surface density profiles inside the cavity, namely rising (Equation \eqref{eq:sigmai-andrews}), flat (Equation \eqref{eq:sigmai-flat}), and declining $\Sigma_{\rm i}$ (Equation \eqref{eq:sigmai-linearneg}). The surface densities for these models are illustrated in Figure~\ref{fig:sigma}. Panel (a) in Figure~\ref{fig:rp-variation} shows the effect of varying the surface density on the image radial profiles. Except for shifting the entire curve up and down, different $\Sigma_{\rm i}$ produce qualitatively very similar images and radial profiles, and all contain the gross features in many SEEDS observations (illustrated by the scaling relation (\ref{eq:seeds})). This is due to both the fact that smooth surface density and scale height profiles yield a smooth scattering surface, and the effect of the PSF. We note that there is some coronagraph edge effect at the inner working angle in the masked images, which is caused by that the part of the disk just outside (but not inside) the mask is convolved with the PSF. This results in a narrow ring of flux deficit just outside the mask. In general, the flux is trustable beyond about one FWHM of the PSF from $\psi_{\rm in}$ ($\sim0.\!\!''05+0.\!\!''15=0.\!\!''2$) \citep[some instrumental effects in observations may also affect the image quality within one FWHM from the mask edge as well]{mut12}.

\subsubsection{The effect of $\delta_{\rm cav,s}$}\label{sec:image-deltacav015}

We investigate the effect of different $\delta_{\rm cav,s}$ with flat $\Sigma_{\rm i}$ (\ref{eq:sigmai-flat}). Panel (b) in Figures~\ref{fig:rp-variation} shows the effect on the radial profile of the convolved images. When deviating from the fiducial model with $\delta_{\rm cav,s}=1$ (i.e. a continuous small dust disk at $R_{\rm cav}$), a bump around the cavity edge and a surface brightness deficit inside the cavity gradually emerge. We quantify this effect by measuring the relative flux deficit at $0.\!\!''2$ ($\frac{2}{3} R_{\rm cav}$) as a function of $\delta_{\rm cav,s}$, the small-dust discontinuity at the cavity edge (subpanel in each plot). For each model we calculate the ratio of the flux at $0.\!\!''2$, $f_{0.2}$, to $f_{0.6}$, the flux at $0.\!\!''6$ ($= 2 R_{\rm cav}$), normalized by $f_{0.2}/f_{0.6}$ in the fiducial, undepleted model. For the model with a $50\%$ discontinuity in $\Sigma_{\rm s}$ at $R_{\rm cav}$ ($\delta_{\rm cav,s}=0.5$), $f_{0.2}/f_{0.6}$ is $\sim$15\% lower than in the fiducial model in the convolved unmasked images, and $\sim$20\% lower in the convolved masked images. The latter is larger because the mask suppresses the halo of the innermost disk, thus the relative flux deficit in the masked image is closer to its intrinsic value, i.e.~the deficit in the raw images. In this sense the masked images are better at constraining the discontinuity of the small dust than the unmasked images. The middle row in Figure~\ref{fig:image-variation} shows the model images for $\delta_{\rm cav,s}=0.3$ (illustrated by the thin solid curve in Figure~\ref{fig:sigma}). The edge of the cavity is quite prominent in the raw image, while it is somewhat smeared out but still visible in the convolved images. For many objects in this IRS/SMA/Subaru cross sample, the SEEDS images are grossly consistent with a continuous small dust disk at the $R_{\rm cav}$, while in some cases a small discontinuity may be tolerated.

For the purpose of comparison with the observations, we now discuss the detectability of a finite surface density discontinuity at the cavity edge, given the sensitivity and noise level of the SEEDS data (the numerical noise level in our simulations is well below the noise level in the observations, see Section~\ref{sec:setup}). In typical SEEDS observations with an integration time of several hundred seconds, the intrinsic Poisson noise of the surface brightness radial profile due to finite photon counts is usually a few tenth of one percent at radius of interests, smaller than the error introduced by the instrument and the data reduction process. \citet{mut12} estimated the local noise level of the surface brightness to be $\sim$10\% at $R=0.\!\!''5$ for the SEEDS SAO 206462 PI images, which should be an upper limit for the noise level in the azimuthally averaged surface brightness (more pixels) and in the inner region of the images (brighter). If this is the typical value for the instrument, then our modeling results indicate that SEEDS surface brightness measurements should be able to put relatively tight constraints on the surface density discontinuity for the small dust at $R_{\rm cav}$. For example, SEEDS should be able to distinguish a disk of small dust continuous at $R_{\rm cav}$ from a disk of small dust with a $50\%$ density drop at $R_{\rm cav}$.

Thus, a lower limit on the small dust depletion factor at the cavity edge can be deduced from detailed modeling for each individual object. This lower limit is likely to be higher than the upper limit from SMA on the depletion factor of the big dust ($\lesssim0.1-0.01$), for objects with relatively smooth radial profiles. If this is confirmed for some objects in which the two limits are both well determined, it means that the density distribution of the small dust needs to somehow {\it decouple} from the big dust at the cavity edge. We will come back to this point in Section~\ref{sec:discussion}.

\subsubsection{The effect of the puffed up inner rim and cavity wall}\label{sec:image-puffingup}

Lastly, we explore the effects of the puffed up inner rim and the cavity wall on the images, by comparing the fiducial model (no puffing up anywhere) to a model with the inner rim puffed up by 100\%, and another model with the cavity wall puffed up by 200\%. Panel (c) in Figure~\ref{fig:rp-variation} shows the effects. Puffing up the inner rim has little effect on the image, while the puffed up wall produces a bump at the cavity edge, similar to the effect of a gap edge. Images for the model with the puffed up wall are shown in Figure~\ref{fig:image-variation} (bottom row), where the wall is prominent in the raw image while been somewhat smeared out but still visible in the convolved images. Scattered light images should be able to constrain the wall for individual objects. Without digging deeply into this issue, we simply note here that the SEEDS images for many objects in this sample are consistent with no or only a small puffed up wall.

\section{The sub-mm properties and images}\label{sec:submm}

\subsection{The sub-mm intensity profile}\label{sec:submm-intensity}

For all models shown in Section~\ref{sec:image}, the disk is optically thin in the vertical direction at 880 $\mu$m ($\tau\sim\Sigma\kappa_\nu\lesssim1$, where $\kappa_\nu$ is the opacity per gram at $\nu$). In this case, the intensity $I_\nu(R)$ at the surface of the disk at a given radius may be expressed as:
\begin{equation}
I_\nu(R)\approx\int^{\infty}_{-\infty}{\left(B_{\nu,T_{\rm b}(z)}\kappa_{\nu,\rm b}\rho_{\rm b}(z)+B_{\nu,T_{\rm s}(z)}\kappa_{\nu,\rm s}\rho_{\rm s}(z)\right)dz}
\label{eq:intensity880}
\end{equation}
where $\nu=341$~GHz at 880 $\mu$m, $B_{\nu,T(z)}$ is the Planck function at $T(z)$ (the temperature at $z$), and $\rho(z)$ is the vertical density distribution ($T$ and $\rho$ depend on $R$ as well). Outside the cavity, the big dust dominates the 880 $\mu$m emission, since the big dust is much more efficient at emitting at $\sim$880 $\mu$m than the small dust ($\kappa_{\nu,\rm b}/\kappa_{\nu, \rm s} \gtrsim 30$ at these wavelengths). Inside the cavity there is {\it no} big dust by our assumption, so the sub-mm emission comes only from the small dust. The thick curves in Figure~\ref{fig:intensity880} show the 880 $\mu$m intensity as a function of radius for the models in Section~\ref{sec:image-sigmai} (i.e. continuous small dust disks with rising, flat, or declining $\Sigma_{\rm i}$). In other words, this is an 880 $\mu$m version of the surface brightness radial profile for the ``raw'' image, without being processed by a synthesized beam dimension (the SMA version of the ``PSF''). The bottom thin dashed curve is for the uniformly heavily depleted cavity model as in Figure~\ref{fig:image-andrews} ($\delta_{\rm s}=\delta_{\rm cav,s}\approx10^{-5}$, rising $\Sigma_{\rm i}$) and no big dust inside the cavity, which represents the sub-mm behavior of the models for most systems in A11.

Since the disk is roughly isothermal in the vertical direction near the mid-plane where most dust lies \citep{chi97}, $T(z)$ may be approximated by the mid-plane temperature $T_{\rm mid}$. The Planck function can be approximated as $B_{\nu,T}\sim2\nu^2 kT/c^2$ at this wavelength due to $h\nu\ll kT$ (880 $\mu$m $\sim$16~K, marginally true in the very outer part of the disk). In addition, since the two dust populations have roughly the same mid-plane temperature, but very different opacities ($\kappa_{\nu,\rm b}\gg\kappa_{\nu,\rm s}$), Equation~\eqref{eq:intensity880} can be simplified to $I_\nu(R)\propto T_{\rm mid}\kappa_{\nu,\rm b}\Sigma_{\rm b}$ for $R>R_{\rm cav}$ and $I_\nu(R)\propto T_{\rm mid}\kappa_{\nu,\rm s}\Sigma_{\rm s}$ for $R<R_{\rm cav}$. For our continuous small disk models with a complete cavity for the big dust, the intensity at 880 $\mu$m drops by $\sim$2.5 orders of magnitude when moving from outside ($I_\nu(R_{\rm cav}+\epsilon)$) to inside ($I_\nu(R_{\rm cav}-\epsilon)$) the cavity edge, due to both the higher opacity of the big dust and the fact that big dust dominates the mass at $R>R_{\rm cav}$. Inside the cavity, the intensity (now exclusively from the small dust) is determined by the factor $T_{\rm mid}\Sigma$. For an irradiated disk $T_{\rm mid}$ increases inwardly, typically as $T_{\rm mid}\propto R^{-1/2}$ \citep{chi97}, while $\Sigma_{\rm i}$ in our models could have various radial dependencies (Equations~\eqref{eq:sigmai-andrews}-\eqref{eq:sigmai-linearneg}). In the flat $\Sigma_{\rm i}$ models (Equation~\eqref{eq:sigmai-flat}), the intensity inside the cavity roughly scales with $R$ as $I_\nu(R)\propto R^{-1/2}$ --- the same as \{$T_{\rm mid}(R)$\} --- and is 1.5 orders of magnitude lower than $I_\nu(R_{\rm cav}+\epsilon)$ in the innermost disk (around the sublimation radius). On the other hand, if there is no depletion of the small dust anywhere inside the cavity (i.e.~the rising $\Sigma_{\rm i}$ with $\delta_{\rm s}=1$ case), the intensity at the center (thick dashed curve) can exceed $I_\nu(R_{\rm cav}+\epsilon)$ by two orders of magnitude.

In A11, for 880 $\mu$m images of models with no big dust inside the cavity, the residual emission near the disk center roughly traces the quantity $T\Sigma_{\rm i}\kappa$. In most cases, A11 found those residuals to be below the noise floor. Due to the sensitivity limit, the constraint on $T\Sigma_{\rm i}\kappa$ inside the cavity is relatively weak; nevertheless, A11 were able to put an upper limit equivalent to $\delta_{\rm b}\lesssim0.01-0.1$ for the mm-sized dust (with exceptions such as LkCa 15). Here we use a mock disk model to mimic this constraint. The top thin dashed curve in Figure~\ref{fig:intensity880} is from a model with uniform depletion factors $\delta_{\rm s}=\delta_{\rm b}=0.01$ for {\it both} dust populations inside the cavity (so the entire disk has the same dust composition everywhere). The result shows that, qualitatively, various models with a continuous small dust disk and a complete cavity for the big dust are all formally below this mock SMA limit, though a quantitative fitting of the visibility curve is needed to constrain $\delta_{\rm s}$ and $\delta_{\rm cav,s}$, in terms of upper limits, on an object by object basis. This may line up with another SED-based constraint on the amount of small dust in the innermost disk, as we will discuss in the Section~\ref{sec:sed}.

\subsection{The 880 $\mu$m images}\label{sec:submm-image}

While the intensity discussion qualitatively demonstrates the sub-mm properties of the disks, Figure~\ref{fig:880um} shows the narrow band images at 880 $\mu$m for two disk models. The top row is from the model which produces Figure~\ref{fig:image-andrews} (also the bottom thin dashed curve in Figure~\ref{fig:intensity880} and the left panel in Figure~\ref{fig:sigma}), which is an A11 style model with a uniformly heavily depleted cavity with rising $\Sigma_{\rm i}$, $\delta_{\rm s}=\delta_{\rm cav,s}\approx10^{-5}$, and no big dust inside the cavity. The bottom row is from the fiducial model in Section~\ref{sec:imageresult} (which produces the top row in Figure~\ref{fig:image-variation}, and the thick solid curve in Figure~\ref{fig:intensity880} and in the left panel in Figure~\ref{fig:sigma}), which has a continuous distribution for the small dust with flat $\Sigma_{\rm i}$ and $\delta_{\rm cav,s}=1$, and no big dust at $R<R_{\rm cav}$ as well. The panels are the raw images from the radiative transfer simulations (left), images convolved by a Gaussian profile with resolution $\sim$$0.\!\!''3$ (middle, to mimic the SMA observations, A11) and $\sim$$0.\!\!''1$ (right, to mimic future ALMA observations (Section~\ref{sec:future}).

Both models reproduce the characteristic features in the SMA images of this transitional disk sample: a bright ring at the cavity edge and a flux deficit inside, agree with the semi-analytical analysis in Section~\ref{sec:submm-intensity}, but {\it very different} NIR scattered light images. The intrinsic reason for this apparent inconsistency is, as we discussed above, that big and small dust dominate the sub-mm and NIR signals in our models, respectively. Thus two disks can have similar images at one of the two wavelengths but very different images at the other, if they share similar spatial distributions for one of the dust populations but not the other.

Lastly, we comment on the effect of big to small dust ratio, which is fixed in this work as 0.85/0.15 to simplify the model (see the discussion of depletion of the small dust in the surface layer of protoplanetary disks, \citealt{dal06}). The scattering comes from the disk surface and is determined by the grazing angle, which only weakly depends on the small dust surface density, if it is continuous and smooth (Section~\ref{sec:image-sigmai}). Changing the mass fraction of the big dust in the outer disk from 0.85 to 0.95 in our $\delta_{\rm cav,s}=1$ models (effectively a factor of 3 drop in surface density of the small dust everywhere) introduces a $\sim$20\% drop in the surface brightness of the scattered light images, but a factor of 3 drop in the cavity 880 $\mu$m intensity  ($I_\nu(R)\propto\Sigma_{\rm s}$ at $R<R_{\rm cav}$). Lastly, we note that since the big-to-small dust sub-mm emission ratio is $\propto\kappa_{\nu,\rm b}\Sigma_{\rm b}/\kappa_{\nu,\rm s}\Sigma_{\rm s}$, the small grains must contain more than 90\% of the total dust mass to dominate the sub-mm emission. 

\section{The transitional-disk-like SED}\label{sec:sed}

In this section, we explore the parameter degeneracy in reproducing the transitional-disk-like SED with their distinctive NIR-MIR dips. SED fitting (particularly of the IRS spectrum) can only provide constraints on the spatial distribution of the small dust within a few or a few tens~AU from the center, and it contains strong degeneracy in the parameter space (A11). Below, we show that disk models with different cavity structures can produce roughly the same SED, containing the transitional disk signature, as long as their innermost parts are modestly depleted (by a factor of $\sim$1000 or so). Except for the specifically mentioned parameters, the other parameters of these models are the same as for the fiducial model in Section~\ref{sec:imageresult}; in particular there is no big dust inside the cavity.

\subsection{The degeneracy in SED fitting}\label{sec:sed-degeneracy}

\begin{figure}[tb]
%\vspace*{-0.5cm}
\begin{center}
\epsscale{0.9} \plotone{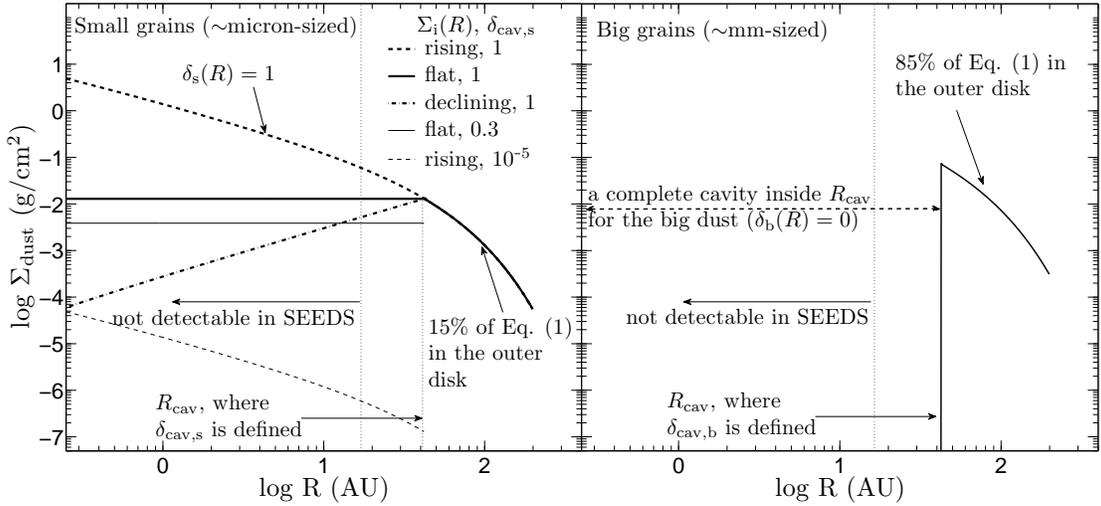}
%\vspace*{-0.5cm}
\end{center} \figcaption{Various surface density profiles of the small dust (up to $\sim$$\mu\rm m$-sized, left) and big dust (up to $\sim$mm-sized, right) in our model (gas-to-dust ratio is assumed to be 100:1 in this work). The big grains contain $85\%$ of the dust mass outside the cavity, and are completely absent inside. The thick curves at $R<R_{\rm cav}$ in the left panel indicate three different $\Sigma_{\rm i}$ (rising, flat, and declining, as Equation~\eqref{eq:sigmai-andrews}-\eqref{eq:sigmai-linearneg}, with names indicating their behavior when moving inward inside the cavity) for the small dust, all continous at the cavity edge ($\delta_{\rm cav,s}=1$). The bottom thin dashed curve illustrates the surface density profile in typical A11 models, which have a uniformly heavily depleted cavity with $\delta_{\rm s}\approx10^{-5}$, and the thin solid curve represents a slight small-grain depletion at the cavity edge ($\delta_{\rm cav,s}=0.3$) with a flat $\Sigma_{\rm i}$. The vertical dotted curves indicate the cavity edge and typical SEEDS inner working angle ($\sim$$0.1-0.15\arcsec$ or $\sim$$15-20$~AU at Taurus). The labels on the axes are for Section~\ref{sec:image} and~\ref{sec:submm}.
\label{fig:sigma}}
\end{figure}

\begin{figure}[tb]
%\vspace*{-0.5cm}
\begin{center}
\epsscale{0.45} \plotone{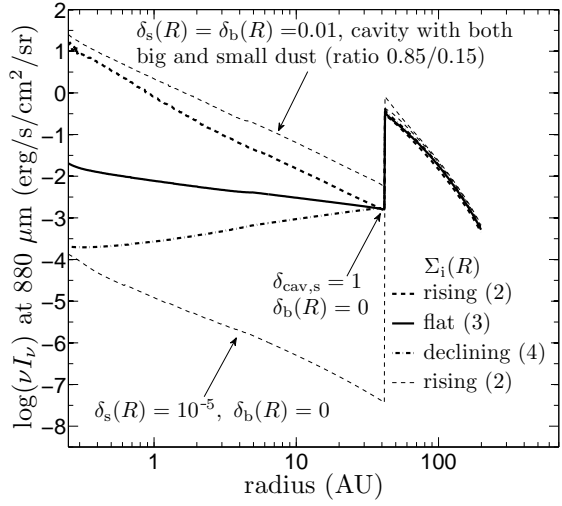}
%\vspace*{-0.5cm}
\end{center} \figcaption{The 880 $\mu$m intensity $I_\nu(R)$ profile (Equation~\eqref{eq:intensity880}) for disk models with various inner disk structures. The model for the top thin dashed curve has a uniformly depleted cavity with $\delta_{\rm s}=\delta_{\rm b}=0.01$ (i.e.~the same dust composition as the outer disk), in order to mimic a ``mock'' SMA upper limit on dust inside the cavity (A11), while all the other models have no big dust ($\sim$mm-sized) inside the cavity. The three thick curves are from models with different $\Sigma_{\rm i}$ and no discontinuity at the cavity edge, corresponding to the same curves in Figure~\ref{fig:sigma} and panel (a) in Figure~\ref{fig:rp-variation}. The bottom thin dashed curve is from a model with a uniformly heavily depleted cavity ($\delta_{\rm s}\approx10^{-5}$). See Section~\ref{sec:submm-intensity} for details of the models. Qualitatively, the disk models with no big dust inside the cavity all produce a deficit in the intensity inside the cavity, since the small dust is very inefficient at sub-mm emission, and all are consistent with the mock upper limit set by the SMA results (A11).
\label{fig:intensity880}}
\end{figure}

\begin{figure}[tb]
%\vspace*{-0.5cm}
\begin{center}
\epsscale{0.45} \plotone{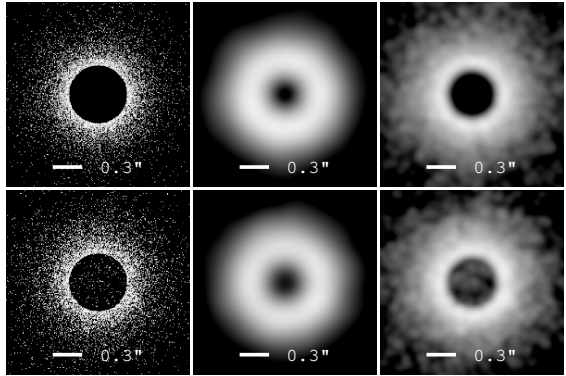}
%\vspace*{-0.5cm}
\end{center} \figcaption{Narrow band 880$\mu$m images for two disk models with no big dust inside the cavity, displayed with logarithmic stretches. Panels are the raw images from the radiative transfer simulations (left), the processed imaged convolved with a Gaussian profile with resolution $0.\!\!''3$ (middle), to mimic SMA observations (A11), and with resolution $0.\!\!''1$ (right), to mimic ALMA observations. The top images are from the uniformly heavily depleted model in Figure~\ref{fig:image-andrews} (rising $\Sigma_{\rm i}$ with $\delta_{\rm s}\approx10^{-5}$, also the thin dashed curve in~\ref{fig:intensity880} and left panel in Figure~\ref{fig:sigma}). The bottom ones are from the same model as the top row in Figure~\ref{fig:image-variation} (also the thick solid curve in Figure~\ref{fig:intensity880} and left panel in Figure~\ref{fig:sigma}), which has a continuous distribution for the small dust ($\delta_{\rm cav,s}=1$) with flat $\Sigma_{\rm i}$ (Equation~\eqref{eq:sigmai-flat}). See Section~\ref{sec:submm-image} for details of the models. The two models produce similar sub-mm images but very different NIR scattered light images (Figure~\ref{fig:image-andrews} and~\ref{fig:image-variation}). The bottom right panel shows that with its exceptional spatial resolution ($\sim$$0.\!\!''1$ or better), ALMA may reveal a sharper edge of the cavity and resolve the small sub-mm emission inside the cavity, which are the keys in a better understanding of the spatial distribution of the grains.
\label{fig:880um}}
\end{figure}

\begin{figure}[tb]
%\vspace*{-0.5cm}
\begin{center}
\epsscale{0.45} \plotone{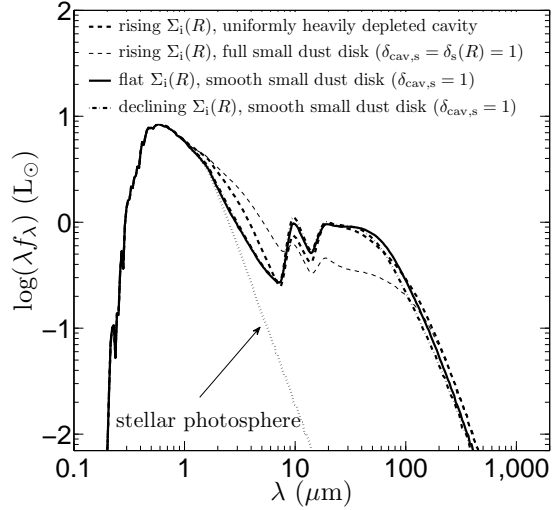}
%\vspace*{-0.5cm}
\end{center} \figcaption{The SED for four disk+cavity models. The uniformly heavily depleted model is motivated by A11, which has an inner cavity with $\delta_{\rm s}\approx10^{-5}$, rising $\Sigma_{\rm i}$, and puffed up inner rim and cavity wall. The full small dust disk model (the thin dashed curve) has otherwise identical properties but with the small dust cavity being uniformly filled up ($\delta_{\rm s}=1$). The other two smooth small dust disk models have $\delta_{\rm cav,s}=1$ (no discontinuity at the cavity edge) and no puffed up inner rim or wall. See Section~\ref{sec:sed-degeneracy} for detailed parameters in these models. The three thick curves show the parameter degeneracy in producing the SED --- diving to roughly the same depth at NIR and coming back to the same level at MIR, as the signature of transitional disks. The full small dust disk model shows that increasing surface density at small radii would eventually wipe out the distinctive deficit in the NIR, and the resulting SED gradually evolves to a full-disk-like SED.
\label{fig:sed}}
\end{figure}

Figure~\ref{fig:sed} shows the SED for four disk models varied based on the fiducial model in Section~\ref{sec:imageresult}. The model for the thick dashed curve has a uniformly heavily depleted cavity with rising $\Sigma_{\rm i}$, $\delta_{\rm s}\approx10^{-5}$, and no big dust inside the cavity (illustrated by the thin dashed curve in the left panel of Figure~\ref{fig:sigma}). The scale height profile has $\beta = 1.15$ and $h/R=0.085$ at 100~AU. The inner rim is puffed up by 100\% and the outer wall is puffed up by 200\%. The inclination is assumed to be $20^\circ$, $R_c=15$~AU, and $R_{\rm cav}=36$~AU. This model is motivated by the A11 disk+cavity structure. The full small dust disk model (the thin dashed curve) has otherwise identical properties but $\delta_{\rm s}=1$ (i.e. completely filled cavity for the small dust). The other two smooth small-dust disk models have much more massive inner disks with $\delta_{\rm cav,s}=1$ (i.e.~a continuous small dust disk) and no puffed up inner rim or cavity wall. The solid curve model has flat $\Sigma_{\rm i}$ (Equation~\eqref{eq:sigmai-flat}, illustrated by the thick solid curve in the left panel of Figure~\ref{fig:sigma}), $\beta=1.33$ and $h/R=0.08$ at 100~AU. The dash-dotted curve model has declining $\Sigma_{\rm i}$ (Equation~\eqref{eq:sigmai-linearneg}, illustrated by the dash-dotted curve in the left panel of Figure~\ref{fig:sigma}), $\beta=1.25$, and $h/R=0.078$ at 100~AU.

The two smooth small dust disk models with flat or declining $\Sigma_{\rm i}$ produce qualitatively similar SED as the uniformly heavily depleted model (in particular, roughly diving to the same depth at NIR, and coming back to the same level at MIR, as the signature of transitional disks), despite the fact that they have very different structures inside the cavity. The minor differences in the strength of the silicate feature and the NIR flux could be reduced by tuning the small dust model and using a specifically designed scale height profile at the innermost part (around the sublimation radius or so). The main reasons for the similarity are:
\begin{enumerate}
\item The depletion factor (or the surface density) at the innermost part (from $R_{\rm sub}$ to $\sim$1~AU or so). While the two smooth small dust disk models differ by $\sim$5 orders of magnitude on the depletion factor (or the surface density) at the cavity edge from the uniformly heavily depleted model, the difference is much smaller at the innermost disk, where most of the NIR-MIR flux is produced. At the innermost disk, the small dust is depleted by $\sim$3 orders of magnitude in the flat $\Sigma_{\rm i}$ model, $\sim$5 orders of magnitude for the declining $\Sigma_{\rm i}$ model, and $\sim$5 orders of magnitude in the uniformly heavily depleted model (with rising $\Sigma_{\rm i}$). On the other hand, the integrated depletion factor $\langle\delta_{\rm s}\rangle$ for the small dust is $\sim$0.3 for the flat $\Sigma_{\rm i}$ model, $\sim$0.2 for the declining $\Sigma_{\rm i}$, and $\sim$10$^{-5}$ for the uniformly heavily depleted model, more in line with $\delta_{\rm cav,s}$, because most of the mass is at the outer part of the cavity. We note that the total amount of small dust is not as important as its spatial distribution inside the cavity, and the amount of dust in the innermost part, in determining the NIR-MIR SED.
\item The scale height of small grains $h_{\rm s}$ at the innermost part. The two smooth small dust disk models are more flared than the uniformly heavily depleted model. While the three have roughly the same scale height outside the cavity, the difference increases inward.  At 1~AU, $h_{\rm s}$ for the uniformly heavily depleted model is 1.7$\times$ that of the flat $\Sigma_{\rm i}$ model and $2.4\times$ that of the declining $\Sigma_{\rm i}$ model.
\item The puffed up inner rim. The inner rim scale height is doubled in the heavily depleted model, which increases the NIR flux and reduces the MIR flux since the puffed up rim receives more stellar radiation and shadows the disk behind it. The puffed up inner rim is removed in the flat or declining $\Sigma_{\rm i}$ models.
\end{enumerate}

The surface density (or the depletion factor) and the scale height at the innermost part are considerably degenerate in producing the NIR to MIR flux in the SED (A11). In general, a disk which has a higher surface density and scale height at the innermost part and a puffed up inner rim intercepts more stellar radiation at small radii, and has more dust exposed at a high temperature, so it produces more NIR flux. On the other hand, the shadowing effect cast by the innermost disk on the outer disk causes less MIR emission \citep{dul04a,dul04b}. In this way, changes in some of these parameters could be largely compensated by the others so that the resulting SED are qualitatively similar.

However, in order to reproduce the characteristic transitional disk SED, the value of the depletion factor inside the cavity cannot be too high. The increasing surface density at small radii would eventually wipe out the distinctive SED deficit, and the resulting SED evolves to a full-disk-like SED, as illustrated by the full small dust disk model in Figure~\ref{fig:sed}. In our experiments with not too flared $\beta$ (comparing with the canonical $\beta\sim1.25-1.3$ in irradiated disk, \citealt{chi97,har98}), we find an upper limit on the order of $10^{-3}$ for the depletion factor in the innermost part in our smooth disk models. We note that this limit depends on the detailed choices of the disk and cavity geometry, such as $R_c$ and $R_{\rm cav}$, and the big-to-small-dust ratio in the outer disk.

\subsection{Discussion of the disk model in producing the SED}\label{sec:sed-discussion}

The inner rim is puffed up in A11 to intercept the starlight and to shadow material at larger radii. At a given radius, the scale height $h$ of the gas disk scales as
\begin{equation}
h\approx c_{\rm s}/\Omega,
\label{eq:h}
\end{equation}
where $\Omega$ is the orbital frequency and $c_{\rm s}$ is the isothermal sound speed in the disk
\begin{equation}
c_{\rm s}\approx\sqrt{kT/\mu},
\label{eq:cs}
\end{equation}
where $T$ is the (mid-plane) temperature and $\mu$ is the average weight of the particles. If the scale heights of the dust and the gas are well coupled (e.g.~for well-mixed-gas-dust models or a constant level of dust settling), tripling the rim scale height (not unusual in A11) means an order of magnitude increase in its temperature. Due to the sudden change of the radial optical depth from $\sim$0 to unity in a narrow transition region directly illuminated by the star, some puffing up may be present, but probably not that significant. In addition, \citet{ise05} pointed out that a realistic puffed up rim has a curved edge (away from the star) instead of a straight vertical edge due to the dependence of $T_{\rm sub}$ on pressure, which further limits the ability of the rim to shadow the outer disk. Based on these reasons, we choose not to have the rim puffed up in our models, though we note that a relatively weak puffing up, as in the uniformly heavily depleted model here, does not make a major difference in the results. Similar idea applies to the puffed up wall as well.

The typical $\beta$ value assumed in A11 in this sample ($\beta\sim1.15$) is small compared with the canonical values for irradiated disk models ($\beta\sim1.25-1.3$, \citealt{chi97,har98}). This leads to that the temperature determined by the input scale height ($T_{\rm input}$, Equation~\eqref{eq:h} and~\eqref{eq:cs}) may increase inwardly too steeply compared with the output mid-plane temperature calculated in the code ($T_{\rm output}$). $T_{\rm input}$ at 100~AU in the three models are close to each other due to their similar scale height there, and all agree with $T_{\rm output}$ ($\sim$30~K) within $20\%$. However, at 1~AU, while $T_{\rm input}$ in our smooth small dust disk models is close to $T_{\rm output}$ ($\sim$220~K, within $30\%$), the input temperature in the uniformly heavily depleted model appears to be too high by a factor of $\sim$3.

%%%%%%%%%%%%%%%%%%%%%%%%%%%%%%%%%%%%%%%%%%%%%%%%%%%%

%%%%%%%%%%%%%%%%%%%%%%%%%%%%%%%%%%%%%%%%%%%%%%%%%%%%

\section{Discussion}\label{sec:discussion}

\subsection{Direct constraints from observations and our generic model}\label{sec:directobsrevations}

First, we review the {\it direct}, model-independent constraints on the disk structure which the three observations --- the infrared SED, SMA sub-mm observations, and SEEDS NIR polarized scattered light imaging --- put on many transitional disks in this cross sample:
\begin{enumerate}
\item IRS reveals a distinctive dip in the spectra around 10 $\mu$m, which indicates that the small dust ($\sim$$\mu$m-sized or so) in the inner part of the disk (from $R_{\rm sub}$ to several~AU) must be moderately depleted. However, due to degeneracy in parameter space, the detailed inner disk structure is model dependent. Models with different cavity depletion factors, $\Sigma_{\rm i}$, and scale heights at the innermost disk could all reproduce the transitional disk signature. The IRS spectra are not very sensitive to the distribution of big dust.
\item The SMA images show a sub-mm central cavity, which indicates that the big dust (mm-sized or so, responsible for the sub-mm emission) is heavily depleted inside the cavity. However, while the observations can effectively constrain the spatial distribution of the big dust outside the cavity, they can place only upper limits on its total amount inside the cavity. SMA observations do not place strong constraints on the distribution of the small dust, though a weak upper limit for the amount of small dust inside the cavity may be determined based on the SMA noise level.
\item SEEDS NIR polarized scattered light images are smooth on large scales, and have no clear signs of a central cavity. The radial profiles of many images increase inwardly all the way from the outer disk to the inner working angle without sudden jumps or changes of slope, indicating that the scattering surfaces and their shapes are smooth and continuous (outside $\psi_{\rm in}$). On the other hand, scattered light images are not very sensitive to the detailed surface density profiles and the total amount of small dust inside the cavity. The NIR images normally do not provide significant constraints on the distribution of the big dust.
\end{enumerate}

In this work, we propose a generic disk model which grossly explains all three observations simultaneously. Previous models in the literature which assume a full outer disk and a uniformly heavily depleted inner cavity can reproduce (1) and (2), but fail at (3), because they also produce a cavity in the scattered light images, which contradicts the new SEEDS results. Through radiative transfer modeling, we find that qualitatively (3) is consistent with a smooth disk of small dust with little discontinuity in both surface density and scale height profile. Table~\ref{tab:models} summarizes the key points in various models and compares their performances in these three observations.

Since we focus on generic disk models only which reproduce the gross features in observations, and we do not try to match the details of specific objects, we more or less freeze many nonessential parameters in Sections~\ref{sec:image}-\ref{sec:sed} which do not qualitatively change the big picture for simplicity. The important ingredients include the dust properties (mostly for the small dust, both the size distribution and the composition), $\Sigma_{\rm s}$, the big-to-small-dust ratio, and $h_{\rm i}$ (both the absolute scale and $\beta$).  NIR scattered light images are able to provide constraints on some of these parameters (particularly $h_{\rm i}$ and $\beta$), due to the dependence of the position and the shape of the disk surface on them. These parameters were not well constrained previously using sub-mm observations and SED due to strong degeneracies (A11).

We note that alternative models for explaining the scattered light images exist, but generally they require additional complications. As one example, if the small dust is not depleted in the outer part of the sub-mm cavity, but is heavily depleted inside a radius smaller than $\psi_{\rm in}$ ($\lesssim15-20$~AU), then it is possible to fit all the three observations, in which case the {\it small dust cavity} does not reveal itself in the scattered light images due to its small size. However, in this case one needs to explain why different dust populations have different cavity sizes. Future scattered light imaging with even smaller $\psi_{\rm in}$ may test this hypothesis. 

As another example, while we achieve a smooth scattering surface by having continuous surface density and scale height profiles for the small dust, it is possible to have the same result with discontinuities in both, but with just the right amount such that the combination of the two yields a scattering surface inside the cavity smoothly joining the outer disk. This may work if the cavity is optically thick (i.e. not heavily depleted), so that a well-defined scattering surface inside the cavity exists. Experiments show that for the models in Section~\ref{sec:imageresult}, uniformly depleting the small dust by a factor of $\sim$1000 and tripling the scale height inside the cavity would roughly make a smooth scattering surface. However, fine tuning is needed to eliminate the visible edge from a small mismatch in the two profiles. Also, the thicker cavity shadows the outer disk, and makes it much dimmer in scattered light (by about one order of magnitude). Lastly, without tuning on the scale height and/or surface density in the innermost part, this model produces too much NIR-MIR flux and too small flux at longer wavelengths in its SED, due to its big scale height at small $R$ and the subsequent shadowing effect.

\subsection{The structure of the cavity in transitional disks}\label{sec:ourfeature}

There are several important conclusions that can be drawn based on our modeling of the transitional disks at different wavelengths.

First, as we discussed in Section~\ref{sec:image-deltacav015}, for some objects the {\it lower} limit for the depletion of the {\it small} dust at the cavity edge (as constrained by the scattered light images) is likely to be above the {\it upper} limit for the {\it big }dust constrained by the SMA, based on the modeling results and the noise level in the two instruments. This essentially means that the small dust has to spatially {\it decouple} from the big dust at the cavity edge. This is the first time that this phenomenon has been associated with a uniform sample in a systematic manner. Detailed modeling of both images for individual objects is needed, particularly in order to determine how sharp the big dust cavity edge is from the sub-mm observations, to pin down the two limits and check if they are really not overlapping. While we defer this to future work, we note that having the big dust the same surface density as the small dust inside the cavity probably cannot reproduce the sub-mm images. Experiments show that even with our declining $\Sigma_{\rm i}$~(\ref{eq:sigmai-linearneg}), a fixed big/small dust ratio and a continuous surface density for both throughout the disk (i.e.~$\delta_{\rm cav,s}=\delta_{\rm cav,b}=1$) produce a sub-mm cavity with a substantially extended edge, and the central flux deficit disappears in the smeared out image. If this is confirmed, it further leads to two possibilities: (a) whatever mechanism responsible for clearing the cavity have different efficiency for the small and big dust, or (b) there are other additional mechanisms which differentiate the small and big dust after the cavity clearing process. At the moment it is not clear which one of the two possibilities is more likely, and both need more thorough investigations.

Second, as we argued in Section~\ref{sec:sed}, in order to reproduce the distinctive NIR deficit in the transitional disk SED, an effective ``upper limit'' of $\delta_{\rm s}$ at the innermost region is required, which in experiments with our disk parameters is on the order of $10^{-3}$. This is far from the {\it lower limit} of $\delta_{\rm s}$ at the cavity edge (close to 1), constrained by the scattered light images. Together, the two limits indicate that the spatial distribution of the small grains is very different inside and outside the cavity --- specifically, $\Sigma_{\rm s}$ tends to be flat or even decrease inwardly inside the cavity.

In addition, this implies that the gas-to-dust ratio needs to increase inwardly, given that most of these objects have non-trivial accretion rates ($\dot{M}\sim10^{-8}-10^{-9}M_{\odot}/$yr, A11). For a steady Shakura \& Sunyaev disk, the accretion rate $\dot{M}$ is related to the gas surface density $\Sigma_{\rm gas}$ as:
\begin{equation}
\frac{\alpha c_{\rm s}^2 \Sigma_{\rm gas}}{\Omega}\approx\frac{\dot{M}}{3\pi},
\label{eq:mdot}
\end{equation}
where $\alpha$ is the Shakura-Sunyaev viscosity parameter, $\Omega$ is the angular velocity of the disk rotation, and $c_{\rm s}$ is given by Equation~\eqref{eq:cs}. At $\sim$0.1~AU, equation~\eqref{eq:mdot} predicts $\Sigma_{\rm gas}\sim$$10^3$~g cm$^{-2}$, assuming a temperature $T\sim10^3$~K, $\dot{M}\sim10^{-8}M_{\odot}/$yr, $\alpha\sim0.01$, and $M\sim M_\odot$ as typical T Tauri values. This is very different from our upper limit of $\Sigma_{\rm gas}\sim$1~g cm$^{-2}$ in the innermost disk, obtained assuming a fixed gas-to-dust ratio of 100 (the flat $\Sigma_{\rm i}$ models in Figure~\ref{fig:sigma}). To simultaneously have large $\Sigma_{\rm gas}$ but small $\Sigma_{\rm dust}$ in the innermost disk, the gas-to-dust ratio needs to increase substantially from the nominal value of 100 (by a factor of $\sim10^3$ in our models, echos with \citealt{zhu11}). This could put constraints on the cavity depletion mechanism or dust growth and settling theory.

At the moment, the mechanism(s) which are responsible for clearing these giant cavities inside transitional disks are not clear (see summary of the current situation in A11). Regarding the applications of our model on this subject, we note two points here. The flat/declining surface density of the small grains inside the cavity in our models is consistent with the grain growth and settling argument, that the small grains in the inner disk are consumed at a faster rate due to higher growth rate there \citep{dul05,gar07,bra08,bir10}. Also, the so-called dust filtration mechanism seems promising for explaining why small dust but not big dust is present inside the cavity \citep{paa06,ric06}, since it could effectively trap the big dust at a pressure maximum in the disk but filter through the small dust. Particularly, combining the two (dust growth and dust filtration), \citet{zhu12} proposed a transitional disk formation model from a theoretical point of view to explain the observations, and their predicted spatial distribution of both dust populations in the entire disk is well consistent with the ones here.

\subsection{Future observations}\label{sec:future}

Imaging is a very powerful tool for constraining the structure of protoplanetary disks and the spatial distribution of both the small and big dust, and there are many ongoing efforts aiming at improving our ability to resolve the disks. In the direction of optical-NIR imaging, updating existing coronagraph and Adaptive Optics (AO) systems, such as the new Coronagraphic Extreme Adaptive Optics (SCExAO) system on Subaru (\citealt{mur10,guy11}, which could raise the Strehl ratio to $\sim$0.9), are expected to achieve better performance and smaller $\psi_{\rm in}$ in the near future.

\begin{figure}[tb]
%\vspace*{-0.5cm}
\begin{center}
\epsscale{0.45} \plotone{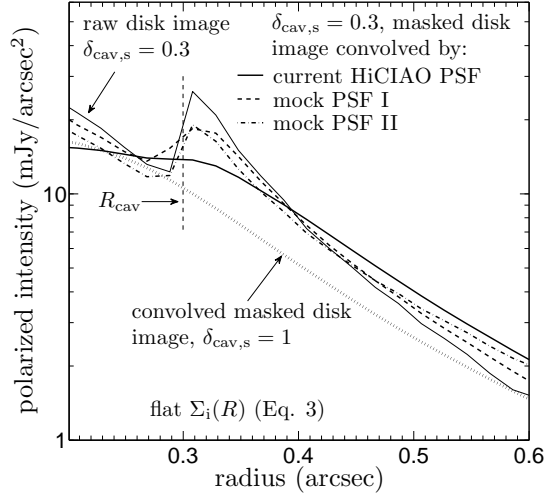}
%\vspace*{-0.5cm}
\end{center} \figcaption{Examples of the power of the optimal performances of the next generation instruments in NIR imaging (such as the SCExAO on Subaru and thirty meter class telescopes). The thick solid, dashed, and dot-dashed curves show the surface brightness radial profile of several $H$ band masked disk images convolved from the {\it same} raw image by several {\it different} PSF (indicated by the legend, see Section~\ref{sec:future} for details). The base model corresponds to the middle row in Figure~\ref{fig:image-variation}, which has a $0.\!\!''3$ radius cavity, flat $\Sigma_{\rm i}$, and $\delta_{\rm cav,s}=0.3$ (a $70\%$ drop in ${\Sigma_{\rm s}}$ at $R_{\rm cav}$). No coronagraph stellar residual (see Section~\ref{sec:psf}) is added in order to isolate the effect of the PSF, and we bin the images into annulus $0.\!\!''02$ in width for better illustration. For comparision, the thin solid line is for the raw (unconvolved) image, and the dotted line is from the corresponding full small dust disk model (i.e.~the top row in Figure~\ref{fig:image-variation}). These examples show that with the ability of next generation instruments expected in the next decade or so, the transition of the spatial distribution of the small dust at the cavity edge could be constrained much better.
\label{fig:nextgenerationpsf}}
\end{figure}

To demonstrates the power of the optimal performances of these next generation instruments in the NIR imaging, Figure~\ref{fig:nextgenerationpsf} shows the surface brightness radial profile of several masked $H$-band disk images convolved from the {\it same} raw image by {\it different} PSF. Except the dotted curve, all the other curves are from the model corresponding to the middle row in Figure~\ref{fig:image-variation}, which has a $0.\!\!''3$ radius cavity, flat $\Sigma_{\rm i}$, and $\delta_{\rm cav,s}=0.3$ (a $70\%$ drop in $\Sigma_{\rm s}$ at $R_{\rm cav}$). We use three PSFs: the current HiCIAO PSF in H band, which could be roughly approximated by a diffraction limited core of an 8-m telescope (resolution $\sim$$0.\!\!''05$) with a Strehl ratio of $\sim$0.4 plus an extended halo; mock PSF I (to mimic SCExAO), which is composed of a diffraction limited core of an 8-m telescope with a Strehl ratio of 0.9, and an extended halo similar in shape (but fainter) as the current HiCIAO PSF; model PSF II (to mimic the next generation thirty-to-forty meter class telescopes), which is composed of a diffraction limited core of a 30-m telescope (resolution $\sim$$0.\!\!''013$) with a Strehl ratio of 0.7, and a similar halo as the previous two. Compared with the full small dust disk case, all the convolved images of the $\delta_{\rm cav,s}=0.3$ model shows a bump at $R_{\rm cav}$ and a relative flux deficit at $R<R_{\rm cav}$. However, from the current HiCIAO PSF to model PSF I and II, the contrast level of the cavity becomes higher and higher, and closer and closer to the raw image (which essentially has an infinite spatial resolution). With these next generation instruments, the transition of the spatial distribution of the small dust at the cavity edge will be better revealed.  

On the other hand, in radio astronomy, the Atacama Large Millimeter Array (ALMA\footnote{http://www.almaobservatory.org/}) is expected to revolutionize the field, with its much better sensitivity level and exceptional spatial resolution ($\sim$$0.\!\!''1$ or better). As examples, the right panels in Figure~\ref{fig:880um} show images convolved by a Gaussian profile with resolution $\sim$$0.\!\!''1$, which mimic the ability of ALMA and show two prominent improvements over the images under the current SMA resolution ($\sim$$0.\!\!''3$, middle panels). First, the edge of the cavity is much sharper in the mock ALMA images. This will make the constraint on the transition of the big dust distribution at the cavity edge much better. Second, while the weak emission signal in the bottom model is overwhelmed by the halo of the outer disk in the mock SMA image, resulting in that the bottom model is nearly indistinguishable from the top model (which essentially produces zero cavity sub-mm emission), the mock ALMA image successfully resolves the signal as an independent component from the outer disk, and separates the two models. This weak emission signal traces the spatial distribution of the dust (both populations) inside the cavity, which is the key in understanding the transitional disk structure.

At this stage, the total number of objects which have been observed by all three survey-scale projects (using IRS/SMA/Subaru) is still small. Increasing the number in this multi-instrument cross sample will help clear the picture. In addition, future observations which produce high spatial resolution images at other wavelengths, such as UV, optical, or other NIR bands (for example using HST, \citealt{gra09}, or the future JWST), or using interferometer (such as the Astrometric and Phase-Referencing Astronomy project on Keck, \citealt{ade07,pot09}, and AMBER system on Very Large Telescope Interferometer, \citealt{pet07,tat07}) should also be able to provide useful constraints on the disk properties.

%%%%%%%%%%%%%%%%%%%%%%%%%%%%%%%%%%%%%%%%%%%%%%%%%%%%

\section{Summary}\label{sec:summary}

We summarize this paper by coming back to the question which we raised at the beginning: what kind of disk structure is consistent with and is able to reproduce the characteristic signatures in all three observations of transitional protoplanetary disks: a high contrast cavity in sub-mm images by SMA, a NIR deficit in SED by Spitzer IRS, and a smooth radial profile in NIR polarized scattered light images by Subaru HiCIAO. We propose one generic solution for this problem, which is feasible but by no means unique. The key points are:

\begin{enumerate}
\item A cavity with a sharp edge in the density distribution of big grains (up to $\sim$mm-sized) and with a depletion factor of at least 0.1-0.01 inside is needed to reproduce the SMA sub-mm images, as pointed out by A11.
\item Right inside the cavity edge ($\sim$15-70~AU), the surface density for the small dust ($\sim$micron-sized and smaller) does not have a big sudden (downward) jump (A small discontinuity may exist). The SEEDS NIR scattered light images, which typically detect the disk at $R\gtrsim$15~AU (the inner working angle in SEEDS), generally require continuous/smooth profiles for the surface density and scale height of the small dust.
\item The small dust in the innermost region (i.e.~within a few~AU, on a scale smaller than measured by SEEDS) has to be moderately depleted in order to produce the transitional-disk-like SED, assuming the disk is not too flared, but the exact depletion factor is uncertain and model dependent.
\end{enumerate}

As we discussed in Section~\ref{sec:ourfeature}, combining all the above points, our model suggests that the spatial distributions of the big and small dust are {\it decoupled} inside the cavity (particularly at the cavity edge). Also, our model argues that the surface density of the small dust inside the cavity is flat or decreases with radius, consistent with the predictions in dust growth models. Combined with the accretion rate measurement of these objects, it further implies that the gas-to-dust ratio increases inwardly inside the cavity of transitional disks.

%%%%%%%%%%%%%%%%%%%%%%%%%%%%%%%%%%%%%%%%%%%%%%%%%%%%

\section*{Acknowledgments}

R.D. thanks Sean Andrews, Nuria Calvet, Eugene Chiang, Bruce Draine, Catherine Espaillat, Elise Furlan, and Jim Stone for useful conversations and help. This work is partially supported by NSF grant AST 0908269 (R. D., Z. Z., and R. R.), AST 1008440 (C. G.), AST 1009314 (J. W.), AST 1009203 (J. C.), NASA grant NNX22SK53G (L. H.), and Sloan Fellowship (R. R.). We thank Pascale Garaud and Doug Lin for organizing the International Summer Institute for Modeling in Astrophysics (ISIMA) at Kavli Institute for Astronomy and Astrophysics, Beijing, which facilitated the discussion of radiative transfer modeling among R. D., L. H., T. M., and Z. Z.. We would also like to thank the anonymous referee for suggestions that improved the quality of the draft.

%%%%%%%%%%%%%%%%%%%%%%%%%%%%%%%%%%%%%%%%%%%%%%%%%%%%

\begin{center}
\begin{table}
\caption{Model Comparison}
\begin{tabular}{|c|c|c|c|}
\hline
 & Our model & Uniformly heavily depleted cavity & Full small dust
\\ \hline
$\Sigma_{\rm i,s}$ & Equation~\eqref{eq:sigmai-flat}-\eqref{eq:sigmai-linearneg} & Equation~\eqref{eq:sigmai-andrews} & Equation~\eqref{eq:sigmai-andrews} \\
$\delta_{\rm s}$ at $R_{\rm cav}$ ($\delta_{\rm cav,s}$) & 1 & $\sim$10$^{-5}$ & 1 \\
$\delta_{\rm s}$ at $R_{\rm sub}$ & $\sim$10$^{-5}-10^{-3}$ & $\sim$10$^{-5}$ & 1 \\
$\Sigma_{\rm i,b}$ (or $\delta_{\rm b}$) & 0  & 0 & 0 \\
$h_{\rm s}$ at $R_{\rm sub}$ & thinner  & thicker & --- \\
puffed up inner rim & no or little  & little to heavy & --- \\ \hline
observation\tablenotemark{a}: SED & Y & Y & N \\
observation: sub-mm & Y  & Y & Y  \\
observation: NIR & Y & N & Y  \\
\hline
\end{tabular}
\tablenotetext{a}{Whether or not these models can reproduce the signatures in various observations: the infrared deficit around 10 microns in SED, the central cavity in sub-mm images, and the smooth radial profile of NIR images.}
\label{tab:models}
\end{table}
\end{center}

\end{document}